\documentclass[aps,prb,preprint,groupedaddress,amsmath,amssymb,showkeys,nofootinbib]{revtex4-2}

\usepackage{graphicx}
\usepackage{dcolumn}
\usepackage{enumitem}
\usepackage{bm}
\usepackage{physics}
\usepackage{tensor}
\usepackage{slashed}
\usepackage[linktocpage=true]{hyperref}
\hypersetup{
    colorlinks,
    linkcolor={blue},
    urlcolor={blue},
    citecolor={blue},
}
\usepackage{color}
\numberwithin{equation}{section}
\renewcommand{\theequation}{\arabic{section}.\arabic{equation}}

\DeclareMathOperator{\Li}{Li}

\newcommand{\del}{\partial}
\newcommand{\nn}{\nonumber}
\newcommand{\mathZ}{\mathcal{Z}}
\newcommand{\mathD}{\mathcal{D}}
\newcommand{\frakh}{\mathfrak{h}}
\newcommand{\frakz}{\mathfrak{z}}
\newcommand{\frakw}{\mathfrak{w}}
\newcommand{\etatot}{\eta_{\mathrm{tot}}}

\newcommand{\nocontentsline}[3]{}
\newcommand{\tocless}[2]{\bgroup\let\addcontentsline=\nocontentsline#1{#2}\egroup}

\def\C60{A$_x$C$_{60}$}

\def\HgCu3{HgCa$_2$Cu$_3$O$_{8+y}$}
\def\HgCu4{HgBa$_2$Ca$_3$Cu$_4$O$_{10+y}$}
\def\TlCu{Tl$_2$Ba$_2$CuO$_{6+\delta}$}
\def\TlCu3{Tl$_2$Ba$_2$Ca$_2$Cu$_3$O$_{10+y}$}
\def\TlCu4{Tl$_2$Ba$_2$Ca$_3$Cu$_4$O$_{12+y}$}

\def\BiCu3{Bi$_2$Sr$_2$Ca$_{2}$Cu$_3$O$_y$}

\def\8LSCO{La$_{1.88}$Sr$_{.12}$CuO$_4$}

\def\110LNSCO{La$_{1.5}$Nd$_{0.4}$Sr$_{0.1}$CuO$_{4}$}
\def\stage4LCO{La$_{2}$CuO$_{4+\delta}$}
\def\Y248{YBa$_2$Cu$_4$O$_8$}

\def\NbSe2{NbSe$_2$}
\def\TaSe2{TaSe$_2$}
\def\TiSe2{TiSe$_2$}
\def\NaCoOH2O{Na$_{0.3}$CoO$_{2y}$H$_2$O}
\def\MgB2{MgB${}_2$}

\def\URu2Si2{URu$_2$Si$_2$}

\def\Ba122{Ba(Fe$_{1-x}$Co$_x$)$_2$As$_2$}

\begin{document}

\title{Interplay of Quantum and Thermal Fluctuations in Two-Dimensional Randomly Pinned Charge Density Waves}

\author{Matthew C. O'Brien}
\email{mco5@illinois.edu}
\author{Eduardo Fradkin}%
\email{efradkin@illinois.edu}
\affiliation{%
Department of Physics and Anthony J. Leggett Institute for Condensed Matter Theory, Grainger College of Engineering, University of Illinois Urbana-Champaign, 1110 West Green Street, Urbana, IL 61801, USA
}%

\date{\today}

\begin{abstract}
The interplay between quantum and thermal fluctuations in the presence of quenched random disorder is a long-standing open theoretical problem which has been made more urgent by advances in modern experimental techniques. The fragility of charge density wave order to impurities makes this problem of particular interest in understanding a host of real materials, including the cuprate high-temperature superconductors. To address this question, we consider the quantum version of an exactly solvable classical model of two-dimensional randomly pinned incommensurate charge density waves first introduced by us in a recent work, and use the large-$N$ technique to obtain the phase diagram and order parameter correlations. Our theory considers quantum and thermal fluctuations and disorder on equal footing by accounting for all effects non-perturbatively, which reveals a novel crossover between under-damped and over-damped dynamics of the fluctuations of the charge density wave order parameter.
\end{abstract}

\maketitle

\clearpage

\tableofcontents

\clearpage

\section{Introduction \label{sec:intro}}

Quasi-two-dimensional charge density wave (CDW) order has been observed in a wide class of materials, such as the lanthanum cuprates \cite{Li2007,Hucker2011,Lee2022}, dichalcogenides \cite{Rossnagel2011,VanWezel2011}, and a variety of quasi-2D systems. In the special case of the cuprates, the strongly layered crystal structure hosts unidirectional charge density wave (CDW) order which is observed in proximity to the $d$-wave superconducting phase \cite{Tranquada1995,Abbamonte2005,Kivelson2003,Ghiringhelli2012,Mesaros2016}, and has been proposed to be intimately linked, or ``intertwined'', with the superconducting order \cite{Berg2009,Fradkin2015}. Recently, developments in experimental techniques such as x-ray scattering \cite{Jang2016,Mitrano2019,Lee2021}, scanning electron microscopy \cite{Mesaros2016,Fujita2014}, and momentum-resolved electron energy loss spectroscopy (M-EELS) \cite{Vig2017,Kogar2017}, have allowed for precise measurements of dynamic charge correlations in CDW materials. These experimental advances highlight the need for theoretical work to elucidate the properties of charge-ordered states, as the interplay of quantum and thermal fluctuations with the random disorder present in materials remains poorly-understood. In a previous work, we introduced a model of a classical unidirectional incommensurate CDW coupled to quenched random disorder which is exactly solvable in a formal limit \cite{OBrien2024}. In this paper, we extend that model to capture the quantum dynamics of the charge density wave.

The order parameter of a unidirectional CDW, whether it is commensurate with the underlying lattice or not, is the Fourier component of the local charge density $\rho(\vb{x},t)$ at the ordering wave vector $\vb{Q}$:
\begin{equation}
    \rho(\vb{x},t)=\rho_0(\vb{x},t)+ \rho_{\vb{Q}}(\vb{x},t) \, e^{i \vb{Q} \cdot \vb{x}}+\rho_{-\vb{Q}}(\vb{x},t) \, e^{-i \vb{Q} \cdot \vb{x}}+\text{higher  harmonics},
    \label{eq:cdw-op}
\end{equation}
where the uniform density $\rho_0(\vb{x},t)$ is a slowly-varying real field and the CDW order parameter $\rho_{\vb{Q}}(\vb{x},t)=\rho_{-\vb{Q}}^*(\vb{x},t)$ is a slowly-varying complex field. A CDW is an ordered phase of an electronic system in which $\langle \rho_{\vb{Q}}(\vb{x},t)\rangle\neq 0$. This state breaks translation invariance and the point group symmetry of the lattice. Charged impurities in materials produce an effectively random electrostatic potential $V_{\rm imp}(\vb{x})$ which couples linearly to the local charge density $\rho(\vb{x},t)$. Consequently, disorder couples linearly to the CDW order parameter and effectively acts as a quenched local random field. For this reason, CDW states are particularly fragile to disorder. In an incommensurate charge density wave (ICDW) the lattice places no constraint on the phase of the CDW order parameter, and hence, $\rho_{\vb{Q}}(\vb{x},t)$ has a global $U(1)$ symmetry of uniform continuous displacements of the charge density profile, which is spontaneously broken in the ordered phase; the associated Goldstone mode is the well-known sliding mode of the ICDW \cite{McMillan1975,Mitrano2019}. In the rest of this work, we will denote the complex ICDW order parameter as $\rho_{\vb{Q}}(\vb{x},t)\equiv \sigma(\vb{x},t)$ and the ordering wave vector label $\vb{Q}$ will be suppressed for notational clarity.

The theory of disorder in CDW materials has a long history, particularly in classical settings. Early work used the random field $XY$ model as an effective field theory and applied perturbative renormalization group techniques to show that the low temperature critical phase of the clean system is destroyed by impurities, among other results \cite{Houghton1981,Cardy1982,Goldschmidt1982}. However, it soon became clear that non-perturbative techniques were crucial for understanding the role of disorder. For example, the exact functional renormalization group (FRG) technique was used to disprove the dimensional reduction hypothesis \cite{Young1977,Fisher1985,Tarjus2004}. FRG is a particularly valuable tool for understanding the properties of systems in which probability distributions of observables display multifractal behavior and long (slowly decaying) tails \cite{Wiese2022}, and has been applied to answer many questions about classical ICDWs \cite{Giamarchi1994,LeDoussal2006,Andreanov2014,Tarjus2020}.

In this work, we consider the interplay of quantum and thermal fluctuations in an ICDW in the presence of quenched random disorder by extending the non-perturbative method we developed in a recent publication to capture the quantum dynamics of the CDW order parameter \cite{OBrien2024}. Our approach uses the large-$N$ technique (see, for example, Refs. \cite{Stanley1968,Amit-book,Zinn-book,fradkin_2021}), which has been applied by other authors to the study of randomly-pinned ICDWs \cite{Nie2014}, and provides more direct access to quantities we are interested in, such as the dynamic charge susceptibility, which are not accessible with other techniques such as the FRG, particularly in the quantum regime. The motivation for our approach is to provide the first example of a large-$N$ theory of disordered ICDWs with a consistent and solvable clean limit. Although it is well-understood that a theory with a $U(1)$ order parameter can undergo a Berezinskii-Kosterlitz-Thouless (BKT) phase transition in two spatial dimensions at finite temperature \cite{Berezinskii1971,Kosterlitz1973}, this fact had not previously been captured in large-$N$ treatments of this problem. This is because the conventional approach to the large-$N$ technique generalizes the $U(1)$ order parameter manifold of the ICDW to, for example, $U(N)$, sacrificing the ability to describe the topological defects (vortices) unique to $U(1)$. To address this challenge, we considered a two-component generalization of the $\mathbb{C}P^N$ model with a global $U(1)$ symmetry between the components which exists for all $N$. It is this global symmetry manifold on which we encoded the CDW order parameter. Using the fact that the $\mathbb{C}P^N$ model is exactly solvable in the large-$N$ limit \cite{DAdda1978,Witten1979,Coleman-1985}, we devised an interaction between the two components that induces ordering and showed that our model was also exactly solvable in the large-$N$ limit. We then demonstrated that in the absence of disorder the model displays a BKT transition in two dimensions. Next, we included quenched random disorder and again solved the model exactly in the $N\to\infty$ limit as a function of the disorder strength and the coupling between the two $\mathbb{C}P^N$ components, finding a complex phase diagram, including a novel weak-to-strong disorder crossover. We also calculated the disorder-averaged order parameter correlation function and determined its parametric dependence on disorder exactly. Throughout that work we paid special attention to ensuring that our results were consistent with the Imry-Ma theorem, which guarantees the absence of spontaneous continuous symmetry breaking in less than four spatial dimensions \cite{Imry1975}.

In this paper we extend this large-$N$ theory to the quantum case in three spacetime dimensions. We begin by solving our theory in the absence of disorder. At zero temperature, we show that our two-component $\mathbb{C}P^N$ model is also exactly solvable in the large-$N$ limit. Unlike in the two-dimensional classical theory discussed in Ref. \cite{OBrien2024}, the quantum CDW transition is first order, at least in the large-$N$ limit. We then add thermal fluctuations due to finite temperature, and are able to obtain an exact expression in the large-$N$ limit for the free energy as a function of temperature. Although the complex structure of the corresponding saddle-point equations precludes closed-form analytic solutions, we are still able to derive the exact coefficients of the Ginzburg-Landau expansion of the free energy in terms of the amplitude of the CDW order parameter. This allows us to derive the full phase diagram of the quantum theory as a function of temperature and coupling between the two $\mathbb{C}P^N$ components, including a tricritical point separating the low temperature first order transition from the high temperature continuous CDW transition found in Ref. \cite{OBrien2024}.

Having mapped out the phase diagram for the clean system, we then include quenched random disorder. Unlike in a classical equilibrium setting, coupling static disorder to dynamical fields produces nonlocal-in-time interactions. To derive the large-$N$ limit of the model we use the well-understood bilocal field approach \cite{Sachdev1993,Cugliandolo1998,Parcollet1999,Sachdev2010,Kitaev2015,Kitaev2018,Scammell2020}, which allows us to calculate the exact self-energy function of the $\mathbb{C}P^N$ components. The novel weak-to-strong disorder crossover found in the classical model now manifests in the quantum theory as a crossover from under-damped to over-damped dynamics. In the large-$N$ limit the model with disorder is only analytically solvable in certain limits, but with a combination of exact results and numerical solutions we are able to obtain the full phase diagram as a function of disorder and coupling between the $\mathbb{C}P^N$ components at zero temperature. We then calculate the CDW correlation function on both sides of the crossover: On the strong disorder side we show that fluctuations of the CDW are also over-damped at low frequencies and that the source of this damping is exclusively scattering from the static disorder. On the weak disorder side, we find the same apparent violation of the Imry-Ma theorem as in the classical model and explain how it is resolved at order $1/N$ in the large-$N$ expansion. Finally, we turn our attention to the disordered problem at finite temperature. Due to the added complexity we restrict our attention to the case of strong disorder, where analytic results can still be obtained. Here we are able to derive the low temperature dependence of the weak-to-strong disorder crossover and the boundary where the amplitude of the CDW order parameter develops. Finally, we calculate the order parameter correlation function in the strong disorder regime and demonstrate how thermal fluctuations allow for dynamic scattering between the $\mathbb{C}P^N$ fields to enhance the damping of the order parameter.

This paper is structured as follows: In Sec. \ref{sec:clean} we introduce the model we will be studying in this paper and summarize relevant parts of our previous work in Ref. \cite{OBrien2024}. In Sec. \ref{sec:zerotempclean} we present the large-$N$ solution of the model at zero temperature, and in Sec. \ref{sec:finitetempclean} we extend this to finite temperature. In Sec. \ref{sec:cleancorrelations} we examine the order parameter correlations in the absence of disorder. In Sec. \ref{sec:disorder} we discuss how to solve the model with quenched disorder. In Sec. \ref{sec:zerotempdisorder} we present the large-$N$ solution of the disordered model at zero temperature and calculate the order parameter correlations. In Sec. \ref{sec:finitetempdisorder} we do the same at finite temperature. Section \ref{sec:disc} presents our conclusions. Reviews of certain relevant material are presented in two appendices. Appendix  \ref{app:nlsm} presents the solution of the quantum nonlinear sigma model with quenched disorder and Appendix \ref{app:cpn} covers the $\mathbb{C}P^N$ model.

\section{Quantum Dynamics of the Clean System \label{sec:clean}}

\subsection{The Model}

In Ref. \cite{OBrien2024}, we described a classical model whose large-$N$ limit allowed us to study the properties of lower-dimensional order parameters. For completeness, we will summarize here the relevant properties of the model introduced in that work. In this work we will discuss the extension of this model to the quantum case.

We consider a theory of two $N$-component complex fields, $\vb*{z}$ and $\vb*{w}$, which is a two-component generalization of the well-known $\mathbb{C}P^N$ model
\begin{equation}
    S = \frac{1}{g} \int \dd^d \vb{x} \dd\tau \, \left( \abs{D^\mu[a] \vb*{z}}^2 + \abs{D^\mu[a] \vb*{w}}^2 - \frac{K}{g} \abs{\vb*{z}^* \cdot \vb*{w}}^2 \right), \kern3em \abs{\vb*{z}}^2 = \abs{\vb*{w}}^2 = 1, \label{eq:multicompaction}
\end{equation}
where both $g$ and $K$ are positive constants, and $a^\mu$ is a fluctuating $U(1)$ gauge field which is minimally coupled to the complex fields through the covariant derivative $D^\mu[a] = \del^\mu + i a^\mu$. Throughout this work, unless otherwise stated, we work in $D = (d+1)$-dimensional Euclidean spacetime (i.e., imaginary time) $x = (\vb{x},\tau)$, and units where the speed of propagation of the complex fields $\vb*{z}$ and $\vb*{w}$ and the gauge field $a^\mu$ is equal to one. The symmetry group of the action is $U(N)\times U(N)$, which has a $U(1)\times U(1)$ subgroup generated by the following transformations:
\begin{subequations} \label{eq:u1transforms}
\begin{alignat}{3}
    &\text{(i) diagonal (local)} \kern3em &&\vb*{z}(x) \longrightarrow e^{i\phi(x)} \vb*{z}(x), \kern3em &&\vb*{w}(x) \longrightarrow e^{i\phi(x)} \vb*{w}(x), \\
    & && a^\mu(x) \longrightarrow a^\mu(x) - \del^\mu \phi(x), && \nn \\
    &\text{(ii) relative (global)} \kern3em &&\vb*{z}(x) \longrightarrow e^{i\theta} \vb*{z}(x), \kern3em &&\vb*{w}(x) \longrightarrow e^{-i\theta} \vb*{w}(x),
\end{alignat}
\end{subequations}
where $\phi(x)$ is a local $U(1)$ gauge transformation and $\theta$ is a global $U(1)$ transformation.
The global $U(1)$ symmetry of incommensurate charge density wave order can then be identified with the relative $U(1)$ symmetry of the model. Observe that the quartic interaction term in the action is invariant under this relative symmetry which can be spontaneously broken.

The quantum statistical partition function for the theory is
\begin{equation}
    \mathcal{Z} = \int \mathD \lambda_1 \mathD \lambda_2 \mathD a^\mu  \mathcal{D} \vb*{z} \mathD \vb*{w} \exp\left(-S - \int \dd^d \vb{x} \dd\tau \left[ \frac{\lambda_1}{g}( \abs{\vb*{z}}^2 - 1) + \frac{\lambda_2}{g}( \abs{\vb*{w}}^2 - 1) \right] \right),
\end{equation}
where $S$ is the action defined in Eq. \eqref{eq:multicompaction}, and $\lambda_1$ and $\lambda_2$ are independent Lagrange multipliers for the unit vector constraints $\abs{\vb*{z}}^2 = \abs{\vb*{w}}^2 = 1$. The $U(1)$ CDW order parameter can then be made manifest through a Hubbard-Stratonovich transformation
\begin{align}
    \exp\left(\int \dd^d \vb{x} \dd\tau\, \frac{K}{g} \abs{\vb*{z}^* \cdot \vb*{w}}^2 \right) &= \int \mathD \sigma \exp\left(-\int \dd^d \vb{x} \dd\tau\, \left[ \frac{g}{K}  \sigma^* \sigma - \sigma \vb*{z}^* \cdot \vb*{w} - \sigma^* \vb*{z} \cdot \vb*{w}^*  \right] \right), \label{eq:HS_trans}
\end{align}
where $\sigma(x)$ plays the role of the order parameter,\footnote{Strictly speaking, identifying the order parameter requires a Legendre transform. However, it only differs from $\sigma(x)$ at higher order in $1/N$ than we consider in this work \cite{Gross1974}.} and transforms according to $\sigma(x) \rightarrow e^{2i\theta}\sigma(x)$ under the relative $U(1)$ symmetry and is neutral under the local gauge symmetry.
The $\mathbb{C}P^N$ model is a well known generalization of the $O(3)$ nonlinear sigma model. Original interest in the one-component model arose in high-energy physics as a model with instanton solutions in $(1+1)$ dimensions for all values of $N$ and with an exactly solvable large-$N$ limit \cite{DAdda1978,Witten1979}. Models of this type have been used in theories of quantum antiferromagnets where the complex field $\vb*{z}$ plays the role of a Schwinger boson representation of the spin degrees of freedom \cite{Sachdev-1999}. In this work, the $\vb*{z}$ and $\vb*{w}$ fields, as well as the gauge field $a^\mu$, are introduced as ancilla or fractionalized degrees of freedom. Here the physical order parameter $\sigma$ is a composite operator of the fields $\vb*{z}$ and $\vb*{w}$, which mediate interactions for the order parameter $\sigma$ in an effective field theory. We will see in the next section that this theory is exactly solvable in the large-$N$ limit. As such, throughout this work we will primarily be interested in the sector of our theory in which the $SU(N)$ symmetry is unbroken and the gauge potential is (logarithmically) confining. However, we anticipate that our construction could also be applied to physical systems with composite orders such as pair density wave superconductors \cite{Berg2009}, or electronic nematics \cite{Fradkin2010,Fernandes-2019}, in which case the $\vb*{z}$ and $\vb*{w}$ fields would be directly identified with parent order parameters.

In the rest of this section, we will describe the properties of this model at zero and finite temperature in the absence of quenched disorder.

\subsection{Large-\texorpdfstring{$N$}{N} Solution at Zero Temperature\label{sec:zerotempclean}}

At zero temperature, the $SU(N)\times SU(N) \subset U(N)\times U(N)$ global symmetry can be spontaneously broken. To see this, we integrate out only the first $(N-1)$ components of the $\vb*{z}$ and $\vb*{w}$ fields to obtain
\begin{subequations} 
\begin{align}
    \mathcal{Z} &= \int \mathcal{D} \lambda_1 \mathcal{D} \lambda_2 \mathcal{D} a^\mu \mathcal{D} \sigma \mathcal{D} z_N \mathcal{D} w_N\, e^{-(N-1)S_{\mathrm{eff}}}, \\
\begin{split}
    S_{\mathrm{eff}} &= \tr \ln \begin{pmatrix} -D_\mu^2[a] + \lambda_1 & -\sigma \\ -\sigma^* & -D_\mu^2[a] + \lambda_2 \end{pmatrix} + \int \dd^d \vb{x} \dd \tau \left[ \frac{\sigma^* \sigma}{K_0} - \frac{\lambda_1 + \lambda_2}{g_0} \right] \\
    &\kern2em + \frac{1}{g_0} \int \dd^d \vb{x} \dd \tau \begin{pmatrix}
        z_N^* & w_N^*
    \end{pmatrix} \begin{pmatrix} -D_\mu^2[a] + \lambda_1 & -\sigma \\ -\sigma^* & -D_\mu^2[a] + \lambda_2 \end{pmatrix} \begin{pmatrix}
        z_N \\ w_N
    \end{pmatrix}, \label{eq:clean_eff_act}
\end{split}
\end{align}
\end{subequations}
where $g = g_0/(N-1)$ and $K = K_0/(N-1)$. In the $N\rightarrow\infty$ limit, mean field theory becomes exact. Using the uniform ansatz $\lambda_1(x) = \lambda_2(x) = m^2$, $\sigma(x) = \rho e^{i\theta}$, $a_\mu(x) = 0$, $z_N(x) = \psi e^{i\theta/2}$ and $w_N(x) = \psi e^{-i\theta/2}$, where $\theta$ is an arbitrary phase of the relative $U(1)$ symmetry, the partition function becomes
\begin{subequations}
\begin{align}
    \mathZ &= e^{-N \mathcal{V} U_{\mathrm{eff}}}, \\
    U_{\mathrm{eff}} &= \int^\Lambda \frac{\dd^d \vb{q}}{(2\pi)^d} \frac{\dd \omega}{2\pi} \ln \left( [\omega^2 + \vb{q}^2 + m^2]^2 - \rho^2 \right) + \frac{\rho^2}{K_0} - \frac{2m^2}{g_0} + 2(m^2 - \rho)\abs{\psi}^2, \label{eq:effpot}
\end{align}
\end{subequations}
where $\mathcal{V}$ is the $D = (d+1)$-dimensional volume of spacetime and $\Lambda$ is an ultraviolet regulator. In $D=3$, the effective potential Eq. \eqref{eq:effpot} is ultraviolet divergent and requires coupling constant renormalization
\begin{equation}
    \frac{1}{g_0} = \frac{1}{g_R} \left(1 + g_R  \int^\Lambda \frac{\dd^d \vb{q}}{(2\pi)^d} \frac{\dd \omega}{2\pi} \frac{1}{\omega^2 + \vb{q}^2 + \mu^2} \right), \label{eq:gRenorm}
\end{equation}
where $\mu$ is the renormalization scale. The regulator $\Lambda$ can then be removed, yielding the renormalized (physical) potential
\begin{equation}
    U_R = - \frac{1}{6\pi} \left[ (m^2 + \rho)^{3/2} + (m^2 - \rho)^{3/2} - 3 \mu m^2 \right] + \frac{\rho^2}{K_0} - \frac{2m^2}{g_R} + 2(m^2 - \rho)\abs{\psi}^2. \label{eq:effpotclean}
\end{equation}
The mean-field parameters are then determined by solutions to the following saddle-point equations:
\begin{subequations} \label{eq:clean_saddles}
\begin{align}
    \sqrt{m^2+\rho } + \sqrt{m^2-\rho } = 2\mu - 8\pi \left(\frac{1}{g_R} -  \abs{\psi}^2 \right), \\
    \sqrt{m^2+\rho } - \sqrt{m^2-\rho } = 8\pi\left(\frac{\rho}{K_0} - \abs{\psi}^2\right), \\
    \psi(m^2 - \rho) = 0 .
\end{align}
\end{subequations}
This system of equations divides the phase diagram into two sectors: (i) $g_R < g_c = 4\pi/\mu$, or $g_R > g_c$ and $K_0 > 8\pi \mu(1 - g_c/g_R)$, where the solutions are
\begin{subequations}
\begin{align}
    \abs{\psi}^2 = \left(\frac{1}{g_R} - \frac{1}{g_c}\right) + \frac{K_0}{32\pi^2}\left(1 + \sqrt{1 - \frac{8\pi \mu(1 - g_c/g_R)}{K_0}}\right), \\
    m^2 = \rho = K_0\left(\frac{1}{g_R} - \frac{1}{g_c}\right) + \frac{K_0^2}{16\pi^2}\left(1 + \sqrt{1 - \frac{8\pi \mu(1 - g_c/g_R)}{K_0}}\right),
\end{align}
\end{subequations}
and (ii) $g_R > g_c$ and $K_0 < 8\pi \mu(1 - g_c/g_R)$, where the solutions are
\begin{equation}
    \psi = 0, \kern3em m^2 = \mu^2\left(1 - \frac{g_{c}}{g_R}\right)^2, \kern3em \rho = 0,
\end{equation}
Therefore, at fixed $g_R > g_c$, there is a zero temperature phase transition at 
\begin{figure}[!t]
    \centering
    \includegraphics[scale=0.7]{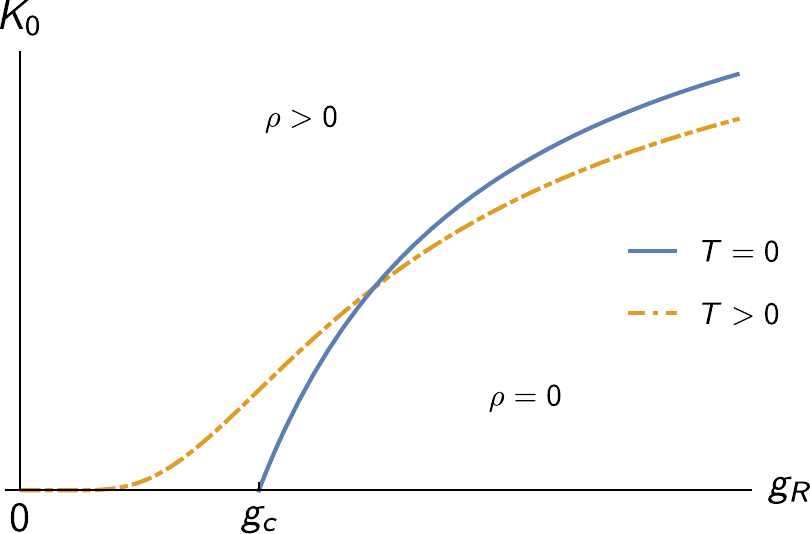}
    \caption{Mean-field ($N\to\infty$) phase diagram as a function of $g_R$, which controls $SU(N)$ ordering ($g_c=4\pi/\mu$ is the $SU(N)$ critical point of the decoupled $\mathbb{C}P^N$ models), and $K_0$, which controls the $U(1)$ CDW ordering. $\rho$ is the large-$N$ expectation value of the CDW order parameter field $\sigma(x)$. The zero temperature phase boundary, Eq. \eqref{eq:zerotempcrit}, is shown in solid blue, while the finite temperature boundary, Eq. \eqref{eq:crit_coupling_tri}  is shown in dashed yellow. At order $1/N$ this finite temperature transition is superseded by a BKT transition, as shown in Ref. \cite{OBrien2024}.}
    \label{fig:phase_diagram_clean1}
\end{figure}
\begin{equation}
    K_{c,0} = 8\pi \mu(1 - g_c/g_R). \label{eq:zerotempcrit}
\end{equation}
The $SU(N)$ order parameter $\abs{\psi}^2$ is continuous at this transition. However, surprisingly, the $U(1)$ condensate $\rho$ has a first order discontinuity of magnitude
\begin{equation} \label{eq:cleanjump}
    \Delta\rho = 2\mu^2 \left(1 - \frac{g_c}{g_R} \right)^2.
\end{equation}
This is in stark contrast to the classical case in two dimensions, where the mean-field $U(1)$ transition was found to be continuous in Ref. \cite{OBrien2024}. The phase diagram is shown in Fig. \ref{fig:phase_diagram_clean1}.

To understand why the transition becomes first order in a higher dimension, we can calculate the scaling dimension of the operator driving the spontaneous breaking of the $U(1)$ symmetry $V(x) =\, : \abs{\vb*{z}^*(x) \cdot \vb*{w}(x)}^2 :$ (where the colons denote normal ordering), at the fixed point $g_R = g_c$ and $K_0 = 0$. At $N=\infty$, the correlator of the local potential $V(x)$ has a free-field  power-law behavior,
\begin{equation}
    \langle V(x) V(y) \rangle \propto \frac{1}{\abs{x - y}^{4(D-2)}},
\end{equation}
which implies its scaling dimension is $\Delta_V=2(D-2)$. Thus, the interaction is relevant for $D < 4$. Since $g_R$ is also a relevant coupling in $D=3$ (while it is marginal in $D=2$), the presence of two relevant operators leads to a runaway renormalization group flow away from the fixed point, which is indicative of a first order transition \cite{Nelson1975,Grest1981,Fradkin1984}.

We note that, unlike the $D=2$ classical theory described in Ref. \cite{OBrien2024}, the ordered phases predicted within mean field theory have genuine long-range order in space-time dimension $D=(2+1)$ at zero temperature; since we are above the lower critical dimension, fluctuations do not play as important a role in the large-$N$ limit. However, it is conceivable that including the fluctuations of the order parameter $\sigma(x)$ with $1/N$ corrections could modify the nature of the phase transition.

\subsection{Large-\texorpdfstring{$N$}{N} Solution at Finite Temperature \label{sec:finitetempclean}}

We now turn on a finite temperature. As usual, the imaginary time direction is compact and $0<\tau<\beta$ where $\beta=1/T$ is the inverse temperature. We know from the Mermin-Wagner theorem that there cannot be an $SU(N)$ symmetry broken phase at finite temperature \cite{Mermin1966}, so we can safely set $\psi(x) = 0$ from hereon. Unlike the $SU(N)$ order, the $U(1)$ order parameter $\sigma(x) = \rho(x) e^{i\theta(x)}$ is a composite of the fundamental $\vb*{z}$ and $\vb*{w}$ fields. As such, it can retain vestigial order with $\rho > 0$ even in the absence of long-range order without contradicting the Mermin-Wagner theorem (see the discussion in Ref. \cite{OBrien2024}). As such, the partition function becomes
\begin{subequations}
\begin{align}
    \mathZ &= e^{-N \beta V U_{\mathrm{eff}}}, \\
    U_{\mathrm{eff}} &= \frac{1}{\beta} \sum_{\omega_n} \int^\Lambda \frac{\dd^d \vb{q}}{(2\pi)^d} \ln \left( [\omega_n^2 + \vb{q}^2 + m^2]^2 - \rho^2 \right) + \frac{\rho^2}{K_0} - \frac{2m^2}{g_0}, \label{eq:effpot_temp}
\end{align}
\end{subequations}
where $V$ is the volume of $d$-dimensional space, $\omega_n = 2\pi n/\beta$ are the bosonic Matsubara frequencies and $U_{\mathrm{eff}}$ has the interpretation of a free energy. One approach to curing the ultraviolet divergence of this potential is with a renormalized coupling that runs with temperature \cite{CastroNeto1993}. However, the following calculations will be more transparent if we subtact all divergences at zero temperature. Following the usual approach (e.g., the Poisson summation formula), we first write the potential/free energy as 
\cite{Chubukov1994},
\begin{align}
\begin{split}
    U_{\mathrm{eff}} = \frac{2}{\beta} \int \frac{\dd^d \vb{q}}{(2\pi)^d} \ln\left(\left[1 - e^{-\beta\sqrt{\vb{q}^2 + m^2 + \rho}}\right]\left[1 - e^{-\beta\sqrt{\vb{q}^2 + m^2 - \rho}} \right] \right) \\
    + \int \frac{\dd^d \vb{q}}{(2\pi)^d} \frac{\dd \omega}{2\pi} \ln \left( [\omega^2 + \vb{q}^2 + m^2]^2 - \rho^2 \right)  + \frac{\rho^2}{K_0} - \frac{2m^2}{g_0},
\end{split}
\end{align}
and after renormalizing the coupling constant at zero temperature as in Eq. \eqref{eq:gRenorm}, we obtain
\begin{align} \label{eq:cleaneffpotrenorm}
\begin{split}
    U_R = - \frac{1}{\pi \beta^2} \left[ \sqrt{m^2 + \rho} \Li_2(e^{-\beta\sqrt{m^2 + \rho}}) +  \sqrt{m^2 - \rho} \Li_2(e^{-\beta\sqrt{m^2 - \rho}})  \right] \\
    -\frac{1}{\pi \beta^3} \left[ \Li_3(e^{-\beta\sqrt{m^2 + \rho}}) + \Li_3(e^{-\beta\sqrt{m^2 - \rho}})  \right] \\ 
   -\frac{1}{6\pi} \left[(m^2 + \rho)^{3/2} + (m^2 - \rho)^{3/2} - 3 \mu m^2 \right]  + \frac{\rho^2}{K_0} - \frac{2m^2}{g_R},
\end{split}
\end{align}
where $\Li_s(z)$ is the polylogarithm function of order $s$ \cite{Gradshteyn-2015}. As a consistency check, we observe that the leading high temperature ($\beta \ll \mu^{-1}$) behavior of the effective potential is
\begin{equation}
    U_R \simeq \frac{m^2}{4\pi \beta} \left[2 - \ln\left(\beta^4[m^4 - \abs{\sigma_0}^2]\right)\right] + \frac{\abs{\sigma_0}}{4\pi \beta} \ln\left(\frac{m^2 - \abs{\sigma_0}}{m^2 + \abs{\sigma_0}}\right) + \frac{\abs{\sigma_0}^2}{K_0} - \frac{2m^2}{g_R} - \frac{2\zeta(3)}{\pi \beta^3},
\end{equation}
where $\zeta(z)$ is the Riemann zeta function. Up to an overall constant and redefinition of the coupling constants $g_R \to \beta g_R$ and $K_0 \to \beta K_0$, this is precisely $\beta^{-1}$ times the classical effective potential we derived in Ref. \cite{OBrien2024}, with renormalization scale $\mu_{\mathrm{classical}} = \beta^{-1}$. The saddle-point equations corresponding to Eq. \eqref{eq:cleaneffpotrenorm} are
\begin{subequations}
\begin{align}
    \sqrt{m^2 -\rho }+\sqrt{m^2 +\rho } + \frac{2}{\beta}\left[\ln \left(1-e^{-\beta  \sqrt{m^2 +\rho }}\right)+\ln \left(1-e^{-\beta  \sqrt{m^2 -\rho }}\right)\right] =2 \mu - \frac{8 \pi }{g_R} , \label{eq:saddle_gap_temp} \\
    \sqrt{m^2 -\rho } - \sqrt{m^2 +\rho } + \frac{2}{\beta}\left[\ln \left(1-e^{-\beta  \sqrt{m^2 +\rho }}\right)-\ln \left(1-e^{-\beta  \sqrt{m^2 -\rho }}\right)\right] =\frac{8 \pi \rho}{K_0}.
\end{align}
\end{subequations}
These equations do not have a closed-form solution for general $m^2$ and $\rho$. However, in the regime where $\rho = 0$ (i.e., small enough $K_0$), we recover the well-known scaling form for the temperature-dependence of the mass parameter $m_0(\beta) = m(\rho=0,\beta)$ \cite{Chubukov1994},
\begin{equation}
    \beta m_0(\beta)= \Psi\left( \beta \mu[1 - g_c/g_R] \right), \kern3em  \Psi(x) = 2 \sinh^{-1}\left(e^{x/2}/2\right) . \label{eq:gap_scaling}
\end{equation}

Since the mean-field $U(1)$ transition is continuous in $D = (2+0)$ and first order in $D = (2+1)$ at zero temperature, we expect that there exists a tricritical point in the phase diagram $(T_{\mathrm{tri}}, K_{\mathrm{tri}})$ which separates the high and low temperature limits, at least for $g_R > g_c$. In the vicinity of such a point $\rho$ must be small, and hence, the effective potential can be expanded as a power series of the Ginzburg-Landau type,
\begin{equation}
    U_R(\rho,\beta)\simeq U_R(0,\beta) + u_1(\beta) \rho^2 + u_2(\beta) \rho^4 + u_3(\beta) \rho^6,
\end{equation}
where symmetry and analyticity demand even powers of $\rho$ in the expansion, and we retain terms up to $\mathcal{O}(\rho^6)$ as is usual for weakly first order phase transitions. To solve for the expansion coefficients, we expand the mass $m(\rho,\beta)$ as a series
\begin{equation}
    [m(\rho,\beta)]^2 \simeq [m_0(\beta)]^2 + m_1(\beta) \rho^2 + m_2(\beta) \rho^4 + m_3(\beta) \rho^6,
\end{equation}
and solve the saddle point equation Eq. \eqref{eq:saddle_gap_temp} term by term. Upon substitution back into the potential, this yields
\begin{subequations}
\begin{align}
    u_1(\beta) &= \frac{1}{K_0} - \frac{\coth(\beta m_0/2)}{8\pi m_0}, \label{eq:ginzburg1}\\
    u_2(\beta) &= \frac{\beta  \csch^3(\beta m_0/2) \sech(\beta m_0/2)}{1536\pi m_0^2} \left( 3 \sinh(\beta m_0) - \beta m_0 \left[ \cosh(\beta m_0) - 2 \right]  \right), \\
\begin{split}
    u_3(\beta) &= -\frac{\beta ^2 \csch^5\left( \beta  m_0/2\right) \sech^3\left(\beta  m_0/2\right) }{2949120 \pi  m_0^{7}} \left(\beta ^2 m_0^2 \left[47 \cosh \left(\beta  m_0\right)-16 \cosh \left(2 \beta  m_0\right) \right. \right. \\
    &\kern3em +\left. \cosh \left(3 \beta  m_0\right)+4\right]+360 \sinh ^2\left(\beta  m_0\right) \\
    &\kern3em - \left.5 \beta  m_0 \left[-51 \sinh \left(\beta  m_0\right)+6 \sinh \left(2 \beta  m_0\right)+\sinh \left(3 \beta  m_0\right)\right]\right) ,
\end{split}
\end{align}
\end{subequations}
where $m_0 = m_0(\beta)$ is given by Eq. \eqref{eq:gap_scaling}. Ginzburg-Landau theory of weakly first order transitions predicts that a tricritical point occurs when $u_1(\beta) = u_2(\beta) = 0$ and $u_3(\beta) > 0$. Solving numerically for $u_2(\beta_{\mathrm{tri}}) = 0$ yields $\beta_{\mathrm{tri}} m_0(\beta_{\mathrm{tri}}) \approx 3.436$; we also confirm that $u_3(\beta_{\mathrm{tri}}) [m_0(\beta_{\mathrm{tri}})]^9 \approx 4\times 10^{-5} > 0$. Using Eq. \eqref{eq:gap_scaling}, this translates to
\begin{equation}
    T_{\mathrm{tri}} \approx 0.2967 \times \mu\left(1 - \frac{g_c}{g_R}\right), \kern3em K_{\mathrm{tri}} \approx 0.9376 \times  K_{c,0} . \label{eq:tricritical}
\end{equation}
In the vicinity of the tricritical point the critical coupling $K_c$ for the formation of a CDW amplitude $\rho > 0$ follows from Ginzburg-Landau theory,
\begin{figure}[!t]
    \centering
    \includegraphics[scale=0.7]{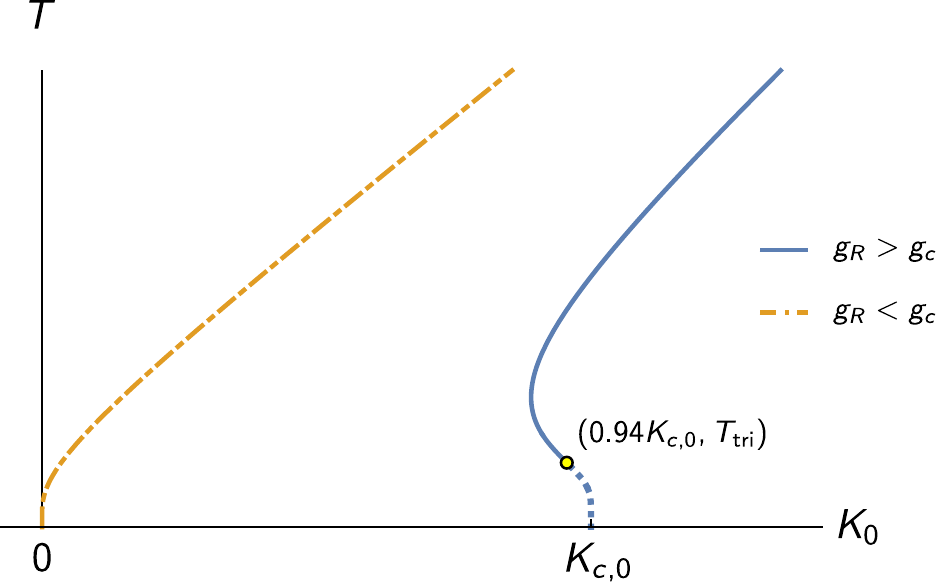}
    \caption{Mean-field ($N\to \infty$) phase diagram as a function of $K_0$, which controls $U(1)$ CDW ordering, and $T$. The boundaries $K_c$ are given by Eq. \eqref{eq:crit_coupling_tri}; $\rho > 0$ for $K_0 \geq K_c$ and $\rho=0$ for $K_0 < K_c$. The boundary for $g_R > g_c$ (the decoupled $\mathbb{C}P^N$ models are $SU(N)$ symmetric at zero temperature) is shown in solid blue, with the tricritical point [see Eq. \eqref{eq:tricritical}] marked as a yellow dot and the first order transitions shown as a dotted line. The boundary for $g_R < g_c$ (broken $SU(N)$ symmetry at zero temperature) is shown in dashed yellow. Both finite temperature transitions are superseded by a BKT transition at order $1/N$. The zero temperature critical point is given by $K_{c,0} = 8\pi \mu(1 - g_c/g_R)$ for $g_R > g_c$.}
    \label{fig:phase_diagram_clean2}
\end{figure}
\begin{equation}
    K_c^{-1} = \begin{cases}
        \dfrac{\coth(\beta m_0/2)}{8\pi m_0} + \dfrac{[u_2(\beta)]^2}{2 u_3(\beta)}, \kern3em & T_* \lesssim T< T_{\mathrm{tri}} , \\[1em]
        \dfrac{\coth(\beta m_0/2)}{8\pi m_0} , \kern3em &  T_{\mathrm{tri}} < T,
    \end{cases} \label{eq:crit_coupling_tri}
\end{equation}
where $\beta_*m_0(\beta_*) \approx 5.372$ is the point at which the $u_3(\beta)$ coefficient changes sign at low temperature. Below this temperature the transition is necessarily beyond the weakly first order Ginzburg-Landau paradigm and the expansion is not strictly valid; the $T\rightarrow 0$ limit of this expression is, however, consistent with the exact $T=0$ solution discussed above. We note that while $u_3(\beta)$ also becomes negative as $\beta \rightarrow 0$, we know from the exact solution of the classical theory that the $\mathcal{O}(\rho^8)$ term in the potential is positive \cite{OBrien2024} (while it is not as $\beta\rightarrow\infty$) and that the transition is continuous. Finally, we note that the first order transition at low temperatures destroys the conventional quantum critical ``fan'' picture of temperature-coupling phase diagrams \cite{Sachdev-1999}. We will see below that the order parameter gap remains finite at $K_c$ when the transition is first order, indicating the suppression of fluctuations.

Importantly, the above analysis of the Ginzburg-Landau expansion only holds for $g_R > g_c$, as can be seen from Eq. \eqref{eq:tricritical}. When $g_R \leq g_c$, $\beta m(\beta) \leq 2\ln((\sqrt{5}+1)/2) \approx 0.96$, which always corresponds to the high temperature continuous phase transition regime. In this case, Eq. \eqref{eq:ginzburg1} predicts a critical coupling $K_c(T) = 4\pi T \exp(\mu(1 - g_c/g_R)/T)$. The phase diagram as function of $g_R$ and $K_0$ at fixed temperature in the continuous transition regime is shown in Fig. \ref{fig:phase_diagram_clean1}. Evidently, the zero temperarture critical point at $g_R = g_c$ and $K_0 = 0$ is rounded by thermal fluctuations, and the critical coupling for $g_R < g_c$ is lifted to $K_c > 0$, though it remains exponentially suppressed. The phase diagram as a function of $K_0$ and $T$ at fixed $g_R$ is shown in Fig. \ref{fig:phase_diagram_clean2}, with the drastically different behavior on either side of $g_c$ evident from the low temperature limits and the existence of the tricritical point when $g_R > g_c$. Since our primary interest is in modeling CDW order, from hereon we will largely restrict our attention to the case $g_R > g_c$ where the decoupled $\mathbb{C}P^N$ components $\vb*{z}$ and $\vb*{w}$ are in the $SU(N)$ symmetric phase at zero temperature.

We emphasize that the phase diagrams derived in this section are mean-field results which are exact only in the $N\to\infty$ limit, and that, as discussed in Ref. \cite{OBrien2024}, the finite temperature phase diagram will be modified to order $1/N$. The classical $(2+0)$-dimensional theory, which corresponds to the high temperature limit of the model considered here, has a Berezinskii-Kosterlitz-Thouless (BKT) phase transition driven by the proliferation of vortices in the phase $\theta(x)$ of the order parameter $\sigma(x)=\rho(x)e^{i\theta(x)}$ \cite{Berezinskii1971,Kosterlitz1973}. At order $1/N$ the mean-field transition is actually a crossover at which the CDW amplitude develops. In contrast, it is known that for continuous transitions in the three-dimensional $XY$ universality class, long-range order and the $U(1)$ amplitude form simultaneously; there is no intermediate regime analogous to the 2D vortex plasma, even though vortex loop excitations are understood to play an important role \cite{Shenoy1989,Williams1999}. In a quantum system with such a 3D $XY$ transition at zero temperature, the long wavelength physics at any finite temperature is always dominated by the 2D BKT renormalization group flow. However, as we have already pointed out, the model we consider in this work does \textit{not} have a continuous transition at zero temperature, so the quantum to classical crossover will necessarily be more complicated,
and the nature of the transition will depend on the actual location of the tricritical point $T_{\mathrm{tri}}$.

\subsection{Order Parameter Correlations \label{sec:cleancorrelations}}

We now consider dynamics of the order parameter $\sigma(x)$, first focusing on the featureless symmetric regime at small $K_0$ where there is no CDW amplitude ($\rho = 0$). In this case it is natural to write the complex order parameter in terms of its real and imaginary parts $\sigma(x) = \sigma^R(x) + i \sigma^I(x)$, in which case the leading quadratic contribution to the effective (Euclidean) action is
\begin{equation}
    S_{\mathrm{eff}} = \frac{N}{2\beta} \sum_{a=R,I} \sum_{\omega_n} \int \frac{\dd^2 \vb{p}}{(2\pi)^2} \sigma^{(a)}(-\vb{p},-\omega_n)  \Pi_\sigma(\vb{p},\omega_n) \sigma^{(a)}(\vb{p},\omega_n), \label{eq:OP_effAct}
\end{equation}
where the kernel is the same for the real and imaginary parts because we are in the symmetric phase,
\begin{equation}
    \Pi_\sigma(\vb{p},\omega_n) = \frac{2}{K_0} - \frac{2}{\beta} \sum_{\omega_{n'}} \int \frac{\dd^2 \vb{p}'}{(2\pi)^2}  G(\vb{p}',\omega_{n'}) G(\vb{p} + \vb{p}',\omega_n + \omega_{n'}) ,
\end{equation}
where
\begin{equation}
  G(\vb{p},\omega_n) = \frac{1}{\omega_n^2 + \vb{p}^2 + m^2}  
\end{equation}
is the Euclidean propagator of the $\vb*{z}$ and $\vb*{w}$ fields in the large-$N$ limit. The real frequency response is then given by the analytic continuation back to real frequency $\omega_n \rightarrow -i \omega$, after which we can expand for small $\vb{p}$ and $\omega$ to find the dynamic susceptibility of the order parameter
\begin{equation}
    \langle \sigma(\vb{p},-i\omega) \sigma^*(-\vb{p},i\omega) \rangle =    \frac{2}{\Pi_\sigma(\vb{p},-i \omega)} \simeq \frac{1}{m_\sigma^2 + \gamma_\sigma \left(\vb{p}^2 - v_\sigma^{-2} \omega^2 \right)},
\end{equation}
where the mass gap $m_\sigma$, stiffness $\gamma_\sigma$ and velocity $v_\sigma$ are
\begin{subequations}
\begin{align}
    m_\sigma^2 = \frac{1}{K_0} - \frac{\coth(\beta m/2)}{8\pi m} \longrightarrow \frac{1}{K_0} - \frac{1}{8\pi m}, \\
    \gamma_\sigma = \frac{[\beta  m+\sinh (\beta  m)] \csch^2\left(\beta m /2\right)}{192 \pi  m^3}  \longrightarrow \frac{1}{96 \pi m^3},\\
    v_\sigma^{-2} = 1 + \frac{2 \beta ^2 m^2 \coth \left(\beta  m/2\right)}{\beta  m+\sinh (\beta  m)}  \longrightarrow 1,
\end{align}
\end{subequations}
where the arrows indicate the zero temperature limits in the regime $g_R > g_c$. Note that the analysis in the previous subsection guarantees that $m_\sigma^2 > 0$ everywhere in the symmetric phase. In fact, in the weakly first order transition regime, Eq. \eqref{eq:crit_coupling_tri} predicts a finite gap $m_\sigma^2 = [u_2(\beta)]^2/u_3(\beta)$ at the transition.

Just as in the large-$N$ solution of the nonlinear sigma model, the order parameter does not display any damping to leading order in $N$, even at finite temperature \cite{Chubukov1994}; that is, the propagation of excitations of the order parameter is ballistic at this order. As we will see below, this stands in stark contrast with the behavior in the presence of quenched disorder.

We note for completeness that the dynamic susceptibility which is, in principle, experimentally accessible corresponds to the physical (and gauge-invariant) composite operator $\mathcal{O} \equiv \vb*{z}\cdot\vb*{w}^*$, and not the Hubbard-Stratonovich field $\sigma$. As noted above, these two quantities are related by a Legendre transform, and the dynamic susceptibility (in real frequency) is
\begin{equation}
    \chi_{\mathcal{O}}(\vb{p},\omega) = -\frac{N}{K_0} + \frac{N}{K_0^2} \langle \sigma(\vb{p},-i\omega) \sigma^*(-\vb{p},i\omega) \rangle = \frac{N}{K_0^2} \frac{2 - K_0 \Pi(\vb{p},-i\omega)}{\Pi(\vb{p},-i\omega)}.
\end{equation}
Since the only quantitative difference is the addition of a constant (contact) term, we see that all salient physical features are contained in the correlation function of the Hubbard-Stratonovich field $\sigma$. As such, we will only consider a description in terms of $\sigma$ for the rest of this work.

Next, we consider the large $K_0$ regime where there is a CDW amplitude ($\rho > 0$). Here it is more appropriate to write the order parameter in its amplitude-phase representation $\sigma(x) = \rho(x) e^{i\theta(x)}$. Far enough away from the transition, the fluctuations of the amplitude mode $\rho$ will be gapped and weak, so we can freeze it out and consider only the fluctuations of the phase field $\theta(x)$, the Goldstone mode of the spontaneously broken $U(1)$ translation symmetry of the incommensurate CDW state. Expanding the effective action Eq. \eqref{eq:clean_eff_act} to quadratic order in the Goldstone mode $\theta(x)$ yields
\begin{equation}
    S_{\mathrm{eff}} = \frac{N}{2\beta} \sum_{\omega_n} \int \frac{\dd^2 \vb{p}}{(2\pi)^2} \theta(-\vb{p},-\omega_n) \Pi_\theta(\vb{p},\omega_n) \theta(\vb{p},\omega_n),
\end{equation}
where the kernel is
\begin{align}
\begin{split}
    \Pi_\theta(\vb{p},\omega_n) = -\frac{2}{\beta} \sum_{\omega_{n'}} \int \frac{\dd^d \vb{p}'}{(2\pi)^d} \left[ \rho^2 G_{zz}(\vb{p}',\omega_{n'}) G_{zz}(\vb{p} + \vb{p}',\omega_n + \omega_{n'}) \right. \\
    - \left. \rho^2 G_{zw}(\vb{p}',\omega_{n'}) G_{zw}(\vb{p} + \vb{p}',\omega_n + \omega_{n'}) - \rho G_{zw}(\vb{p}',\omega_{n'}) \right],
\end{split}
\end{align}
and the $\vb*{z}$ and $\vb*{w}$ propagator is
\begin{equation}
    G^{-1}(\vb{p},\omega_n) = \begin{pmatrix} \omega_n^2 + \vb{p}^2 + m^2 & -\rho \\ -\rho & \omega_n^2 + \vb{p}^2 + m^2 \end{pmatrix} .
\end{equation}
After analytic continuation $\omega_n \rightarrow -i \omega$ and expanding for small $\vb{p}$ and $\omega$, the phase field propagator has the manifestly gapless form
\begin{equation}
    \langle \theta(\vb{p},-i\omega) \theta(-\vb{p},i\omega) \rangle = \frac{1}{\Pi_\theta(\vb{p},-i\omega)} \simeq \frac{1}{\gamma_\theta \vb{p}^2 - \chi_\theta \omega^2},
\end{equation}
where the phase stiffness $\gamma_\theta$ and susceptibility $\chi_\theta$ have unwieldy exact expressions, but simplify in the limit of small $\rho$,
\begin{subequations}
\begin{align}
    \gamma_\theta &\simeq  \frac{ \csch^2(\beta m_0/2) [\beta  m_0+\sinh (\beta  m_0)]}{96 \pi  m_0^3} \rho ^2, \\
    \chi_\theta &\simeq \gamma_\theta\left(1 - \frac{\beta^2 m_0^2 \coth(\beta m_0/2)}{\beta m_0 + \sinh(\beta m_0)} \right),
\end{align}
\end{subequations}
where $m_0 = m_0(\beta)$ is the mass in the absence of a condensate $\rho$. Note that, as we pointed out in Ref. \cite{OBrien2024}, thermal fluctuations of the phase mode destroy long range CDW order. The zero temperature limit is more subtle, as we know that $m^2 = \rho$ in the ordered phase. Therefore, the correct result must be obtained by taking $\beta \rightarrow\infty$ in the exact expressions, followed by setting $m^2 = \rho$, yielding the phase stiffness and susceptibility
\begin{equation}
    \gamma_\theta = \chi_\theta = \frac{\sqrt{\rho}}{12\sqrt{2}\pi}.
\end{equation}
As expected, the velocity of the phase mode $v_\theta = \sqrt{\gamma_\theta/\chi_\theta} = 1$ since the system is Lorentz invariant at zero temperature.

\section{Quantum Dynamics with Quenched Disorder \label{sec:disorder}}

We now turn to the main focus of this work: the role of quenched random disorder. As in Ref. \cite{OBrien2024}, we consider disorder with random fields $\frakz^a(\vb{x})$, $\frakw^a(\vb{x})$, and $\frakh^a(\vb{x})$, transforming under the adjoint representation of $U(N)$. These random fields are coupled to the two $\mathbb{C}P^N$ components $\vb*{z}(\vb{x},\tau)$ and $\vb*{w}(\vb{x},\tau)$ in the following manner,
\begin{equation}
\begin{split}
    \mathcal{L}_{\mathrm{dis}} &= \mathfrak{z}^a(\vb{x}) z^*_\alpha(\vb{x},\tau) \gamma^a_{\alpha\beta} z_\beta(\vb{x},\tau) + \mathfrak{w}^a(\vb{x}) w^*_\alpha(\vb{x},\tau) \gamma^a_{\alpha\beta} w_\beta(\vb{x},\tau) \\
    &\kern2em + \frakh^a(\vb{x}) z^*_\alpha(\vb{x},\tau) \gamma^a_{\alpha\beta} w_\beta(\vb{x},\tau) + \frakh^{a*}(\vb{x}) w^*_\alpha(\vb{x},\tau) \gamma^a_{\alpha\beta} z_\beta(\vb{x},\tau), \label{eq:disorder_coupling}
\end{split}
\end{equation}
and have configurations drawn from the locally Gaussian distributions
\begin{equation}
\begin{split}
    &\kern-2em \overline{\mathfrak{z}^a(\vb{x})} = \overline{\mathfrak{w}^a(\vb{x})} = 0, \kern 2em \overline{\frakz^a(\vb{x}) \frakz^b(\vb{y})} = \overline{\frakw^a(\vb{x}) \frakw^b(\vb{y})} = \eta_1^2 \delta^{ab} \delta^{(d)}(\vb{x}-\vb{y}) , \\
    &\kern-2em \overline{\frakh^a(\vb{x})} = 0, \kern2em \overline{\frakh^{a}(\vb{x}) \frakh^b(\vb{y})}  = 0, \kern2em \overline{\frakh^{a*}(\vb{x}) \frakh^b(\vb{y})} = \eta_2^2 \delta^{ab} \delta^{(d)}(\vb{x}-\vb{y}) , \label{eq:disorder_ensemble}
\end{split}
\end{equation}
where $\frakz^a(\vb{x})$ and $\frakw^a(\vb{x})$ are $N^2$-component real vectors, $\frakh^a(\vb{x})$ is an $N^2$-component complex vector, $\gamma^a_{\alpha\beta}$ are the generators of $U(N)$ satisfying $\gamma^a_{\alpha\beta} \gamma^a_{\mu\nu} = N \delta_{\alpha\nu} \delta_{\beta\mu}$, and overlines denote averaging over disorder configurations. For a fixed realization of the disorder fields, each term in Eq. \eqref{eq:disorder_coupling} breaks the $SU(N) \times SU(N)$ global symmetry and is gauge invariant under the local diagonal $U(1)$ symmetry. However, only $\frakh^a(\vb{x})$ breaks the global relative $U(1)$ symmetry of the order parameter. All symmetries are unbroken on average within the ensemble of Eq. \eqref{eq:disorder_ensemble}. We know from Ref. \cite{OBrien2024} that there is a disorder-driven crossover which manifests in replica trick calculations as spontaneous symmetry breaking of the ``replicated'' $U(1)^n$ symmetries (not to be confused with the replica permutation symmetry). In the absence of cross-correlations between $\frakz^a(\vb{x})$ and $\frakw^a(\vb{x})$ (which are allowed by symmetry), the symmetry is spontaneously broken down to the replica-diagonal (i.e., physical) $U(1)$ subgroup. However, if $\overline{\frakz^a(\vb{x})\frakw^b(\vb{y})} \neq 0$, the replica-diagonal global $U(1)$ subgroup can also be spontaneously broken. This is reminiscent of so-called fluctuating order which is induced by random disorder \cite{Kivelson2003}. However, it is understood that the apparent long-range order which appears in that context is an artifact of mean-field theory and the underlying physics is inherently glassy. Such glassy and replica permutation symmetry-breaking physics is beyond the scope of this work, so for our present purposes we only consider the case in which $\overline{\frakz^a(\vb{x})\frakw^b(\vb{y})} = 0$.

To eliminate dependence on any specific realization of the disorder, we use the replica trick to perform disorder averages. The average of the replicated partition function is
\begin{align}
    \overline{\mathZ^n} &= \int \mathD \mathfrak{z} \mathD \mathfrak{w} \mathD \mathfrak{h}\,  \exp\left(- \int \dd^d \vb{x} \left[ \frac{\mathfrak{z}^2 + \mathfrak{w}^2}{2\eta_1^2} + \frac{\abs{\frakh}^2}{\eta_2^2} \right] \right) \mathZ[\frakz,\frakw,\frakh]^n \nn \\ 
\begin{split}
    &= \int \mathcal{D} \lambda_{1,j} \mathcal{D} \lambda_{2,j}  \mathcal{D} a^\mu_j \mathcal{D} \vb*{z}_j \mathcal{D} \vb*{w}_j \prod_{j=1}^n e^{-S_j} \\
    &\times  \exp\left(  \frac{N}{2} \int \dd^d \vb{x} \dd \tau_1 \dd\tau_2 \sum_{i,j=1}^n \left[\eta_1^2  \left( \abs{\vb*{z}^*_i(\tau_1) \cdot \vb*{z}_j(\tau_2)}^2 + \abs{\vb*{w}^*_i(\tau_1) \cdot \vb*{w}_j(\tau_2)}^2 \right) \right.\right. \\
    &\kern6em + \left.\left. 2\eta_2^2 (\vb*{z}^*_i(\tau_1) \cdot \vb*{z}_j(\tau_2)) (\vb*{w}_i(\tau_1) \cdot \vb*{w}^*_j(\tau_2)) \right] \vphantom{\frac{N}{2} \int \dd^d \vb{x} \sum_{i,j=1}^n} \right), \label{eq:disavgedact}
\end{split}
\end{align}
where $S_j$ are the replicas of the original action Eq. \eqref{eq:multicompaction}, and $\vb*{z}_j(\tau) = \vb*{z}_j(\vb{x},\tau)$ and $\vb*{w}_j(\tau) = \vb*{w}_j(\vb{x},\tau)$. The replica-replica interactions induced by the disorder average have important differences compared to those in the classical counterpart to this model studied in Ref. \cite{OBrien2024}: The inclusion of quantum dynamics leads to interactions with a so-called bilocal structure; the static quenched disorder couples fields at different times. In particular, the replica-diagonal terms are no longer trivial, as the unit vector constraint only applies at equal times. We will see that these terms play an important role in governing the order parameter dynamics. Just as the quartic replica-replica interactions place the classical analogue of this model in contrast with the classical $(2+0)$-dimensional $O(N)$ nonlinear sigma model (NLSM), the quartic bilocal interactions are drastically different to those in the quantum $(2+1)$-dimensional NLSM (see Appendix \ref{app:nlsm}). In the NLSM, it is evident that the disorder only couples to the zero frequency component of the $O(N)$ order parameter. In Eq. \eqref{eq:disavgedact}, we see that imaginary time averages are over bilinears $z^*_{\alpha,i}(\vb{x},\tau) w_{\beta,i}(\vb{x},\tau)$, allowing for contributions from finite frequency modes.

To proceed with the large-$N$ technique, we decouple the quartic replica-replica interactions with a Hubbard-Stratonovich transformation. At first, it might be tempting to decouple the interactions in the channel of time-averaged bilinears to mimic the quantum NLSM. However, these bilinears are not $SU(N)$ singlets, and therefore, are not conducive to applying the large-$N$ limit. It is well-understood (see, for example, Refs. \cite{Sachdev1993,Cugliandolo1998,Scammell2020,Parcollet1999,Sachdev2010,Kitaev2015,Kitaev2018}) that the solution is to introduce bilocal fields $\zeta_{ij}(\vb{x},\tau_1,\tau_2) \sim \vb*{z}_i(\vb{x},\tau_1)\cdot \vb*{z}^*_j(\vb{x},\tau_2)$ and $\kappa_{ij}(\vb{x},\tau_1,\tau_2) \sim \vb*{w}_i(\vb{x},\tau_1)\cdot \vb*{w}^*_j(\vb{x},\tau_2)$, so that the interactions in Eq. \eqref{eq:disavgedact} become
\begin{align}
    &\exp\left(  \frac{N}{2} \int \dd^d \vb{x} \dd \tau_1 \dd\tau_2 \sum_{i,j=1}^n \left[ \eta_1^2  \left( \abs{\vb*{z}^*_i(\tau_1) \cdot \vb*{z}_j(\tau_2)}^2 + \abs{\vb*{w}^*_i(\tau_1) \cdot \vb*{w}_j(\tau_2)}^2 \right) \right.\right. \nn \\
    &\kern5em + \left.\left. 2\eta_2^2 (\vb*{z}^*_i(\tau_1) \cdot \vb*{z}_j(\tau_2)) (\vb*{w}_i(\tau_1) \cdot \vb*{w}^*_j(\tau_2)) \right] \vphantom{\frac{N}{2} \int \dd^d \vb{x} \sum_{i,j=1}^n} \right) \nn \\
\begin{split}
    &= \int \mathD \zeta_{ij} \mathD \kappa_{ij} \, \exp\left(-\frac{1}{2N} \sum_{i, j=1}^n  \int \dd^d \vb{x} \dd \tau_i \dd \tau_j  \left[ \eta_1^2 (\zeta_{ij} \zeta_{ji} + \kappa_{ij} \kappa_{ji}) +2 \eta_2^2 \kappa_{ij} \zeta_{ji} \right]\right) \\
    &\times \exp\left( \sum_{i, j=1}^n \int \dd^d \vb{x} \dd \tau_i \dd \tau_j  \left[ \vb*{z}^*_i \cdot \vb*{z}_j (\eta_1^2  \zeta_{ij} + \eta_2^2 \kappa_{ij}) + \vb*{w}^*_i \cdot \vb*{w}_j (\eta_1^2  \kappa_{ij} + \eta_2^2 \zeta_{ij}) \vphantom{\Tilde{\Delta}^2 \vb*{z}_i^* \cdot \vb*{w}_j \kappa_{ij}}  \right]  \right) ,
\end{split}
\end{align}
where, for example, $\vb*{z}_i = \vb*{z}_i(\vb{x},\tau_i)$, and so on. To respect the unit vector constaints on the $\vb*{z}_i(\vb{x},\tau)$ and $\vb*{w}_i(\vb{x},\tau)$, the bilocal fields must satisfy $\zeta_{ii}(\vb{x},\tau,\tau) = \kappa_{ii}(\vb{x},\tau,\tau) = 0$ (this is not a trace condition; no summation over repeated indices). Therefore, after integrating out the $U(N)$ vector fields, we obtain
\begin{align}
\begin{split}
    S_{\mathrm{eff}}/N = \Tr \ln \Bigg[ \delta(\tau_1-\tau_2) \mathrm{diag}_r\begin{pmatrix}
    -D_\mu^2[a] + \lambda_1 & - \sigma \\
    -\sigma^* & -D_\mu^2[a] + \lambda_2
    \end{pmatrix}  - \begin{pmatrix} \tilde{\eta}_1^2 \hat{\zeta} + \tilde{\eta}_2^2 \hat{\kappa} & 0 \\ 0 & \tilde{\eta}_1^2 \hat{\kappa} + \tilde{\eta}_2^2 \hat{\zeta} \end{pmatrix} \Bigg] \\
    + \tr \int \dd^d \vb{x} \dd \tau \left[ \frac{\mathrm{diag}_r(\abs{\sigma}^2)}{K_0}  - \frac{\mathrm{diag}_r(\lambda_1 + \lambda_2)}{g_0}  \right] + \tr \int \dd^d \vb{x} \dd \tau_1 \dd \tau_2 \left[ \frac{\tilde{\eta}_1^2}{2} (\hat{\zeta}^2 + \hat{\kappa}^2) + \tilde{\eta}_2^2  \hat{\zeta} \hat{\kappa} \right] ,  \label{eq:disorderedEffAct}
\end{split}
\end{align}
where we have rescaled $(\zeta,\kappa) \rightarrow g_0 (\zeta,\kappa)$, and defined $g = g_0/N$, $K = K_0/N$, and $\eta_{1,2} = \tilde{\eta}_{1,2}/g_0$ to obtain a well-defined large-$N$ limit. $\mathrm{Tr}(\,\cdot\,)$ includes the functional operator trace as well as the trace over replica indices, $\mathrm{diag}_r(\,\cdot\,)$ denotes a matrix which is diagonal in replica indices, and $\hat{\zeta}$ and $\hat{\kappa}$ are the matrices with elements $\zeta_{ij}$ and $\kappa_{ij}$, respectively. As in Ref. \cite{OBrien2024}, we restrict our attention to the case where $\tilde{\eta}_1 > \tilde{\eta}_2$ to avoid any spontaneous breaking of the replica permutation symmetry. The disorder Hubbard-Stratonovich fields transform as tensors under the replicated $U(1)^n$ symmetries:
\begin{subequations}
\begin{alignat}{2}
    &\text{(i) diagonal} \kern3em && \zeta_{jk}(\vb{x},\tau_1,\tau_2) \longrightarrow e^{i(\phi_j(\vb{x},\tau_1) - \phi_k(\vb{x},\tau_2))} \zeta_{jk}(\vb{x},\tau_1,\tau_2), \\
    & &&\kappa_{jk}(\vb{x},\tau_1,\tau_2) \longrightarrow e^{i(\phi_j(\vb{x},\tau_1) - \phi_k(\vb{x},\tau_2))} \kappa_{jk}(\vb{x},\tau_1,\tau_2),  \nn \\
    &\text{(ii) relative}  && \zeta_{jk}(\vb{x},\tau_1,\tau_2) \longrightarrow e^{i(\theta_j - \theta_k)} \zeta_{jk}(\vb{x},\tau_1,\tau_2), \\
    & && \kappa_{jk}(\vb{x},\tau_1,\tau_2) \longrightarrow e^{-i(\theta_j - \theta_k)} \kappa_{jk}(\vb{x},\tau_1,\tau_2) . \nn
\end{alignat}
\end{subequations}
In terms of the replicas, a physical symmetry which is unbroken by any fixed configuration of the disorder has an unbroken corresponding replicated symmetry (e.g., the local diagonal $U(1)$ symmetry has an unbroken $U(1)^n$ replicated symmetry group), whereas a symmetry which is only unbroken on average in the ensemble of disorder configurations will have a replicated symmetry group which is explicitly broken down to the replica-diagonal subgroup (e.g., the global relative $U(1)$ symmetry of the CDW order parameter). If the Hubbard-Stratonovich fields $\zeta_{jk}$ and $\kappa_{jk}$ condense, then the replicated $U(1)^n$ symmetries will also be spontaneously broken.\footnote{Since the local diagonal $U(1)^n$ symmetry is gauged, it cannot technically be spontaneously broken. However, fluctuations of the collective modes at order $1/N$ will restore the symmetry via the Higgs mechanism.}

To obtain the effective potential, we make the replica permutation-symmetric and uniform ansatz for the local fields $\lambda_{1,j}(x) = \lambda_{2,j}(x) = m^2$, $\sigma_j(x) = \rho e^{i\theta}$, $a^\mu_j(x) = 0$, while the most general ansatz for the bilocal fields is
\begin{align}
    \zeta_{ij}(\vb{x},\tau_1,\tau_2) &= \kappa_{ij}(\vb{x},\tau_1,\tau_2) = \frac{1}{\Tilde{\eta}_1^2 + \Tilde{\eta}_2^2}[\kappa_0(\tau_1 - \tau_2) -\delta_{ij} \kappa_1(\tau_1 - \tau_2)] ,
    \label{eq:kappas}
\end{align}
where $\kappa_0(\tau)$ and $\kappa_1(\tau)$ are functions only of time, and the choice of sign is arbitrary and made for future notational simplicity. Just like the variational parameters $m^2$ and $\rho$, the functions $\kappa_{0}(\tau)$ and  $\kappa_{1}(\tau)$ are to be determined from the saddle-point equations. Making these substitutions and taking the replica limit $n\rightarrow 0$ yields the disorder-averaged effective potential
\begin{align}
\begin{split}
    \overline{U_{\mathrm{eff}}} = \int \frac{\dd^d \vb{q}}{(2\pi)^d} \frac{\dd \omega}{2\pi} \left[ \ln \left( \left[\omega^2 + \vb{q}^2 + m^2 + \tilde{\kappa}_1(\omega) \right]^2 - \rho^2  \right)  - \frac{2 \tilde{\kappa}_0(\omega)\left[\omega^2 + \vb{q}^2 + m^2 + \tilde{\kappa}_1(\omega)\right]}{\left[\omega^2 + \vb{q}^2 + m^2 + \tilde{\kappa}_1(\omega) \right]^2 - \rho^2}  \right] \\
    +\int \dd\tau \left( \frac{[\kappa_1(\tau)]^2 - 2 \kappa_1(\tau) \kappa_0(\tau)}{\eta_{\mathrm{tot}}^2} \right) + \left( \frac{\rho^2}{K_0} - \frac{2 m^2}{g_0} \right) + 2 \alpha  \frac{\kappa_1(0) - \kappa_0(0)}{\eta_{\mathrm{tot}}^{2}} , \label{eq:disorderEffPot}
\end{split}
\end{align}
where $\tilde{\kappa}_{0,1}(\omega)$ are the Fourier transformed functions, $\eta_{\mathrm{tot}}^{2} = \tilde{\eta}_1^2 + \tilde{\eta}_2^2$, and $\alpha$ is the Lagrange multiplier enforcing the constraint $\zeta_{ii}(\vb{x},\tau,\tau) = \kappa_{ii}(\vb{x},\tau,\tau) = 0$. In Eq. \eqref{eq:disorderEffPot} $\kappa_0(0)$ and $\kappa_1(0)$ denote the equal-time values of the functions defined in Eq. \eqref{eq:kappas}. Note that we must have $\tilde{\kappa}_{0,1}(\omega) > 0$ in order for the action to be positive definite. In the following subsections, we will present the solution of this model at zero and then at finite temperature, and explore the order parameter correlations in each case. The solution of the simpler conventional quantum $\mathbb{C}P^N$ model with quenched disorder is instructive, so we also encourage the reader to refer to Appendix \ref{app:cpn}.

\subsection{Zero Temperature \label{sec:zerotempdisorder}}

\subsubsection{Large-\texorpdfstring{$N$}{N} Solution}

The $N\to\infty$ solution of the theory is obtained by solving the saddle point equations for the effective potential Eq. \eqref{eq:disorderEffPot},
\begin{subequations} \label{eq:disorderedsaddles}
\begin{align}
    \int \frac{\dd^d \vb{q}}{(2\pi)^d} \frac{\dd \omega}{2\pi} \left[  \frac{\varepsilon_{\omega,\vb{q}}^2 }{\varepsilon_{\omega,\vb{q}}^4 - \rho^2} + \frac{  \tilde{\kappa}_0(\omega) (\varepsilon_{\omega,\vb{q}}^4 +\rho^2)}{( \varepsilon_{\omega,\vb{q}}^4 - \rho^2 )^2}   \right] = \frac{1}{g_0}, \label{eq:saddleg0}\\
    \int \frac{\dd^d \vb{q}}{(2\pi)^d} \frac{\dd \omega}{2\pi} \left[ \frac{\rho}{\varepsilon_{\omega,\vb{q}}^4 - \rho^2} + \frac{ 2 \rho \tilde{\kappa}_0(\omega)\varepsilon_{\omega,\vb{q}}^2 }{( \varepsilon_{\omega,\vb{q}}^4 - \rho^2 )^2}  \right] = \frac{\rho}{K_0},  \label{eq:saddleK0} \\
    \int \frac{\dd^d \vb{q}}{(2\pi)^d} \left[  \frac{\varepsilon_{\omega,\vb{q}}^2}{\varepsilon_{\omega,\vb{q}}^4 - \rho^2} + \frac{\tilde{\kappa}_0(\omega)  (\varepsilon_{\omega,\vb{q}}^4 +\rho^2)}{( \varepsilon_{\omega,\vb{q}}^4 - \rho^2 )^2}   \right] = - \frac{\Tilde{\kappa}_1(\omega) - \tilde{\kappa}_0(\omega) + \alpha}{\eta_{\mathrm{tot}}^2} , \label{eq:saddlekappa1}\allowdisplaybreaks\\
    \int \frac{\dd^d \vb{q}}{(2\pi)^d} \frac{\varepsilon_{\omega,\vb{q}}^2}{\varepsilon_{\omega,\vb{q}}^4 - \rho^2} = - \frac{\Tilde{\kappa}_1(\omega) + \alpha}{\eta_{\mathrm{tot}}^2} , \label{eq:saddlekappa2}\\
    \kappa_0(0) - \kappa_1(0) = 0, \label{eq:saddlealpha}
\end{align}
\end{subequations}
where $\varepsilon_{\omega,\vb{q}}^2 = \omega^2 + \vb{q}^2 + m^2 + \tilde{\kappa}_1(\omega)$ and the last equation is the equal time constraint imposed by $\alpha$. This system of five equations determines the five parameters and functions $m^2$, $\rho$, $\tilde{\kappa}_0(\omega)$, $\tilde{\kappa}_1(\omega)$ and $\alpha$.

First, we must determine the functions $\tilde{\kappa}_{0}(\omega)$ and $\tilde{\kappa}_{1}(\omega)$ in terms of the other parameters of the theory from Eqs. \eqref{eq:saddlekappa1} and \eqref{eq:saddlekappa2}. We begin by eliminating the Lagrange multiplier $\alpha$ to obtain the expressions
\begin{subequations}
\begin{align}
    \tilde{\kappa}_0(\omega) \int \frac{\dd^d \vb{q}}{(2\pi)^d} \frac{  \left[\omega^2 + \vb{q}^2 + M^2 + \bar{\kappa}_1(\omega) \right]^2 +\rho^2}{\left( \left[\omega^2 + \vb{q}^2 + M^2 + \bar{\kappa}_1(\omega) \right]^2 - \rho^2 \right)^2} = \frac{\tilde{\kappa}_0(\omega)}{\eta_{\mathrm{tot}}^2} , \label{eq:kappa2Eq} \\
    \int \frac{\dd^d \vb{q}}{(2\pi)^d} \left[ \frac{\omega^2 + \vb{q}^2 + M^2 + \bar{\kappa}_1(\omega)}{\left[\omega^2 + \vb{q}^2 + M^2 + \bar{\kappa}_1(\omega) \right]^2 - \rho^2} - \frac{\vb{q}^2 + M^2}{\left(\vb{q}^2 + M^2 \right)^2 - \rho^2} \right] = - \frac{\bar{\kappa}_1(\omega)}{\eta_{\mathrm{tot}}^2}, \label{eq:kappa1Eq}
\end{align}
\end{subequations}
where $M^2 = m^2 + \tilde{\kappa}_1(0)$ and $\bar{\kappa}_1 = \Tilde{\kappa}_1(\omega) - \Tilde{\kappa}(0)$. The subtraction at zero frequency is necessary to make the saddle-point equations finite (see, for example, Ref. \cite{Scammell2020}), and hence, the physical mass is actually $M$; this approach is more natural than performing renormalization of the unphysical parameters $m^2$, $\tilde{\kappa}_1(0)$ and $\alpha$. Next, since $\eta_{\mathrm{tot}}$ is independent of frequency, Eq. \eqref{eq:kappa2Eq} implies that if $\tilde{\kappa}_0(\omega)$ is non-zero, it must have the form
\begin{equation}
    \tilde{\kappa}_0(\omega) = \kappa_0 (2\pi) \delta(\omega),
\end{equation}
where $\kappa_0$ is a constant which we will see corresponds to an effective static response strength, in which case Eq. \eqref{eq:kappa2Eq} becomes
\begin{equation}
    \frac{\kappa_0 M^2}{4\pi(M^4 - \rho^2)} = \frac{\kappa_0}{\eta_{\mathrm{tot}}^2} .
\end{equation}
This equation has two solutions:
\begin{equation}
    \kappa_0 = 0, \kern3em \text{or} \kern3em \kappa_0 \neq 0\kern1em \text{and} \kern1em M^2 = \frac{\eta_{\mathrm{tot}}^2}{8\pi} + \sqrt{\left(\frac{\eta_{\mathrm{tot}}^2}{8\pi}\right)^2 + \rho^2}. \label{eq:disorderpin}
\end{equation}
Depending on whether $\kappa_0 = 0$ or $\kappa_0 \neq 0$, Eq. \eqref{eq:saddleg0} will determine the value of $M^2$ in the former case and $\kappa_0$ in the latter. Next, performing the integral in Eq. \eqref{eq:kappa1Eq} yields the implicit equation for $\bar{\kappa}_1(\omega)$,
\begin{equation}
    \frac{1}{8\pi} \ln\left(\frac{[\omega^2 + M^2 + \bar{\kappa}_1(\omega)]^2 - \rho^2}{M^4 - \rho^2}\right) = \frac{\bar{\kappa}_1(\omega)}{\eta_{\mathrm{tot}}^2} .
\end{equation}
For general values of $\rho$, this equation does not have any explicit solution in terms of known special functions. However, when $\rho = 0$, the mean-field theory is identical to the conventional $\mathbb{C}P^N$ model, and we can learn a great deal from the exact solution which exists in that case; see Appendix \ref{app:cpn} for details. To proceed with the general case, the saddle-point equation can be solved asymptotically in the limit $\omega \rightarrow 0$, and we find very different behavior depending on whether $\kappa_0$ is zero or non-zero:
\begin{subequations} \label{eq:self-energy}
\begin{align}
    \bar{\kappa}_1(\omega) \simeq \left[\frac{4\pi (M^4 - \rho^2)}{\eta_{\mathrm{tot}}^2 M^2} - 1 \right]^{-1} \omega^2 + \mathcal{O}(\omega^4), \kern3em \kappa_0 = 0,\\
    \bar{\kappa}_1(\omega)  \simeq \frac{\eta_{\mathrm{tot}}}{\sqrt{4\pi}} \left( 1 + \frac{1}{\sqrt{1 + (8\pi \rho/\eta_{\mathrm{tot}}^2)^2}}\right)^{1/2} \abs{\omega} + \mathcal{O}(\omega^2), \kern3em \kappa_0 \neq 0.
\end{align}
\end{subequations}
Note that in all cases $M^2 \geq \rho$ and $M^2 > \eta_{\mathrm{tot}}^2/4\pi$. Therefore, we see that the ``strong disorder'' regime where the static response strength $\kappa_0$ condenses also leads to over-damped dynamics of the $SU(N)$ fields; we will see in subsequent sections that the order parameter dynamics will also be over-damped in this regime. On the other hand, in the limit $\abs{\omega} \rightarrow \infty$, we have
\begin{equation}
    \bar{\kappa}_1(\omega) \simeq \frac{\eta_{\mathrm{tot}}^2}{8\pi} \ln\left( \frac{\omega^4}{M^4 - \rho^2}\right),
\end{equation}
in both regimes. Since $\bar{\kappa}_1(\omega)$ is manifestly not a square-integrable function, one might think that the relation between $\bar{\kappa}_1(\omega)$ and $\kappa_1(\tau)$ is pathological, and that the constraint $\kappa_1(\tau=0) = \kappa_0(\tau=0) \equiv \kappa_0$ cannot be satisfied. However, because of the subtraction at zero frequency, we emphasize that $\bar{\kappa}_1(\omega)$ is \textit{not} the Fourier transform of $\kappa_1(\tau)$; this is the source of the cumbersome renormalization which we avoided by performing the subtraction.

Given the form of $\bar{\kappa}_1(\omega)$, we see that the saddle-point equations Eqs. \eqref{eq:saddleg0} and \eqref{eq:saddleK0} can be made finite with the same coupling constant renormalization as in the absence of disorder,
\begin{equation}
    \frac{1}{g_0} = \frac{1}{g_R} \left(1 + g_R  \int^\Lambda \frac{\dd^d \vb{q}}{(2\pi)^d} \frac{\dd \omega}{2\pi} \frac{1}{\omega^2 + \vb{q}^2 + \mu^2} \right),
\end{equation}
where $\mu$ is the renormalization scale, yielding the expressions
\begin{subequations} \label{eq:dirtysaddleszerotemp}
\begin{align}
    \frac{1}{8\pi} \int \frac{\dd \omega}{2\pi} \ln\left(  \frac{\omega^4}{[\omega^2 + M^2 + \bar{\kappa}_1(\omega)]^2 - \rho^2} \right)  = \frac{1}{g_R} - \frac{\mu}{4\pi} - \frac{\kappa_0}{\eta_{\mathrm{tot}}^2}, \label{eq:gappsaddle} \\
    \frac{1}{4\pi} \int \frac{\dd \omega}{2\pi} \tanh^{-1} \left( \frac{\rho}{\omega^2 + M^2 + \Bar{\kappa}_1(\omega)} \right) = \frac{\rho}{K_0} - \frac{\rho \kappa_0}{ \eta_{\mathrm{tot}}^2 M^2} , \label{eq:condensatesaddle}
\end{align}
\end{subequations}
where $\bar{\kappa}_1(\omega)$ is understood in this case to be a known function which depends on the parameters $M^2$, $\rho$ and $\eta_{\mathrm{tot}}$. Together with the condition in Eq. \eqref{eq:disorderpin}, these saddle-point equations determine $M^2$, $\rho$ and $\kappa_0$, and hence, the structure of the phase diagram. Unfortunately, the complicated frequency-dependence of $\bar{\kappa}_1(\omega)$ makes obtaining exact analytic solutions of the saddle-point equations impossible in most cases. 

However, a number of exact results (up to numerical evaluation of dimensionless constants) can be obtained when the disorder is strong. First, it follows from Eqs. \eqref{eq:disorderpin} and \eqref{eq:gappsaddle}, that in the regime where $\rho = 0$, the static response strength as a function of $\etatot$ is
\begin{equation} \label{eq:disorderamplitude1}
    \kappa_0(\eta_{\mathrm{tot}}) = \begin{cases}
        0, \kern3em & \eta_{\mathrm{tot}} \leq \eta_{c,0},\\
        \eta_{\mathrm{tot}}^2 \left( g_R^{-1} - g_{c}^{-1} \right) + \mathsf{c}_1 \eta_{\mathrm{tot}}^3, \kern3em & \eta_{\mathrm{tot}} > \eta_{c,0},
    \end{cases}
\end{equation}
where $g_{c} = 4\pi/\mu$ is the critical coupling of the clean $\mathbb{C}P^N$ model, $\mathsf{c}_1 \approx 0.03941$, and
\begin{equation} \label{eq:critdisorder}
    \eta_{c,0} = \begin{cases}
        0, \kern3em & g_R \leq g_{c},\\
        \mathsf{c}_1^{-1} \left(g_{c}^{-1} - g_R^{-1} \right), \kern3em & g_R > g_{c} ,
    \end{cases}
\end{equation}
is the critical disorder strength in the absence of a CDW condensate; see Appendix \ref{app:cpn} for details of the calculation. This result implies that when the clean system is in its symmetry broken phase, that is, when the spectrum contains gapless excitations, any infinitesimal amount of quenched disorder will be felt strongly. On the other hand, just as in the 2D classical theory, when the clean system is in its gapped phase, there is a crossover between weakly- and strongly-disordered behaviors. Since we are interested in the $U(1)$ CDW order which is a vestige of the ``ancilla'' $\mathbb{C}P^N$ parent fields $\vb*{z}$ and $\vb*{w}$ we will only consider the case where $g_R > g_c$ for the rest of this paper.

\begin{figure}
    \centering
    \includegraphics[scale=0.5]{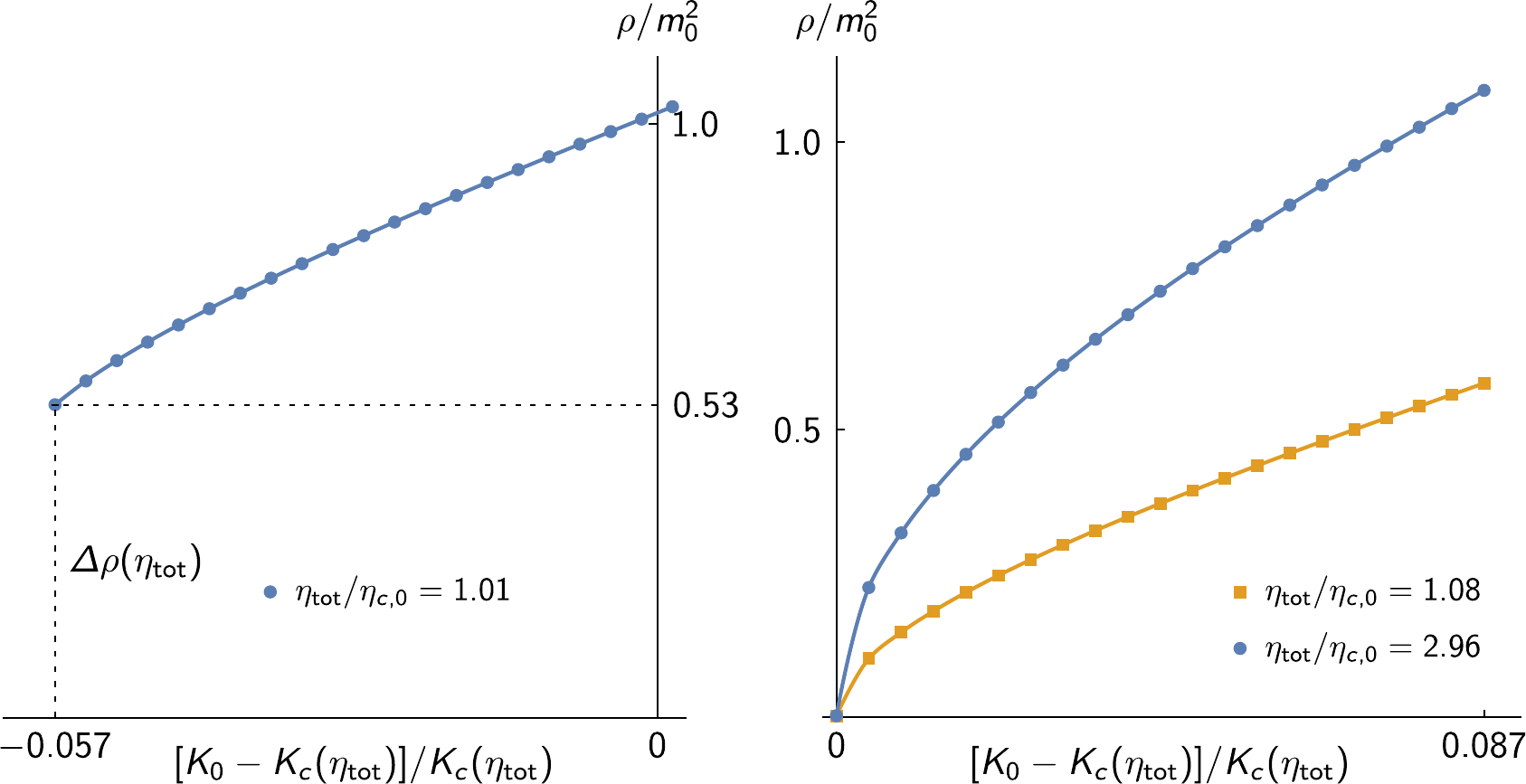}
    \caption{CDW order parameter amplitude $\rho$ as a function of the coupling $K_0$ for $U(1)$ CDW order in the strong disorder regime where the static response strength $\kappa_0 > 0$ and with $g_R/g_c=2$ (the decoupled $\mathbb{C}P^N$ models are $SU(N)$ symmetric at zero temperature and disorder). Points are data determined from numerical solution of Eqs. \eqref{eq:dirtysaddleszerotemp} and lines are guides for the eye. For disorder strength $\etatot$ very close to the disorder-driven crossover point $\eta_{c,0}$, given by Eq. \eqref{eq:critdisorder}, (left panel) the transition is first order with jump $\Delta\rho(\etatot) \approx 0.53 m_0^2$ and the actual critical point at $ K_0  \approx 0.943 \times K_c(\etatot)$ preempts the continuous transition. For large $\etatot$ (right panel) the transition is continuous and the critical $K_0$ coincides with the prediction $K_c(\etatot)$ in Eq. \eqref{eq:dirtycritKzero}.}
    \label{fig:condensatezero}
\end{figure}

\begin{figure}
    \centering
    \includegraphics[scale=0.55]{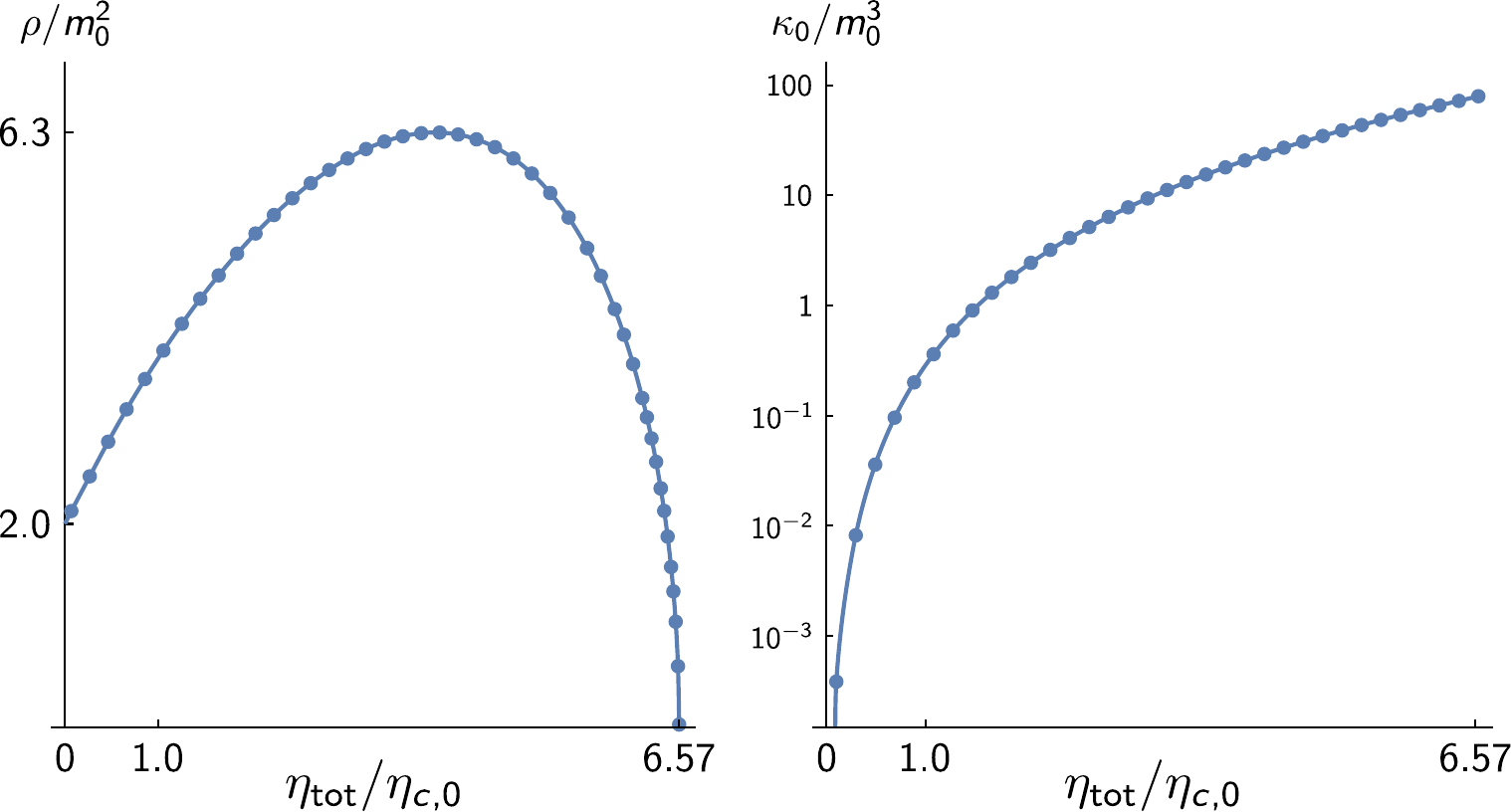}
    \caption{CDW order parameter amplitude $\rho$ (left) and static response strength $\kappa_0$ (right; note the log scale on the vertical axis) as a function of the $SU(N)$ disorder strength $\etatot$ for $g_R/g_c = 2$ (the decoupled $\mathbb{C}P^N$ models are $SU(N)$ symmetric at zero temperature and disorder) and $K_0 = K_{c,0} = 8\pi\mu(1-g_c/g_R)$ the critical coupling of the clean system. Vertical axes are normalized to units of the clean zero temperature gap $m_0 = \mu(1 - g_c/g_R)$. Points are data determined from numerical solution of Eqs. \eqref{eq:dirtysaddleszerotemp} and lines are guides for the eye. $\Delta\rho = 2m_0^2$ is the first order jump at $K_{c,0}$ in the clean system given by Eq. \eqref{eq:cleanjump}. $\eta_{c,0}$ is given by Eq. \eqref{eq:critdisorder}.}
    \label{fig:disorderdependence}
\end{figure}

Next, we consider the regime with $\rho > 0$ and $\kappa_0 > 0$. The CDW condensate $\rho$ as a function of $K_0$ and $\etatot$ can be determined by solving the coupled system in Eqs. \eqref{eq:dirtysaddleszerotemp} numerically, and we find two different behaviors, as shown in Fig. \ref{fig:condensatezero}: For $\eta_{\mathrm{tot}}$ not too large, we observe a first order transition, with a jump in $\rho$ at the critical point. On the other hand, for large enough $\eta_{\mathrm{tot}}$ we observe a continuous transition, with the order parameter $\rho$ approaching zero as $K_0$ approaches a critical value $K_c(\etatot)$. Analogously to the effect of temperature in the clean system discussed in the previous section, the two limiting cases together imply the existence of a tricritical point $(\eta_{\mathrm{tri}}, K_{\mathrm{tri}})$ in the zero temperature phase diagram. This softening of a first order transition into a continuous one is a well-known property of quenched disorder \cite{Aizenman1989,Goswami2008,Hrahsheh2012,Aizenman2012}. While some systems have been observed to display tricritical points \cite{Chatelain2001,Egilmez2008}, it is possible that the first order transition discussed in this section may be an artifact of the $N\to\infty$ limit and could be softened by fluctuations at order $1/N$. With those caveats in mind, in the limit where the transition is continuous, we can actually derive an exact expression for the phase boundary. In this case, Eq. \eqref{eq:condensatesaddle} predicts a critical coupling $K_c(\eta_{\mathrm{tot}})$,
\begin{equation} \label{eq:dirtycritKzero}
    \frac{1}{K_c(\eta_{\mathrm{tot}})} = \frac{\kappa_0(\eta_{\mathrm{tot}})}{\eta_{\mathrm{tot}}^2 M^2} + \frac{1}{4\pi} \int \frac{\dd \omega}{2\pi} \frac{1}{\omega^2 + M^2 + \bar{\kappa}_1(\omega)} = \frac{4\pi}{\eta_{\mathrm{tot}}^2}\left(\frac{1}{g_R} - \frac{1}{g_{c,0}}\right) + \frac{4\pi \mathsf{c}_1 + \mathsf{c}_2}{\eta_{\mathrm{tot}}},
\end{equation}
where $\mathsf{c}_2 \approx 0.1066$, and $\kappa_0(\eta_{\mathrm{tot}})$ is given by Eq. \eqref{eq:disorderamplitude1}. Since the transition at the tricritical point is continuous, $K_{\mathrm{tri}} = K_c(\eta_{\mathrm{tri}})$.

\begin{figure}[!t]
    \centering
    \includegraphics[scale=0.75]{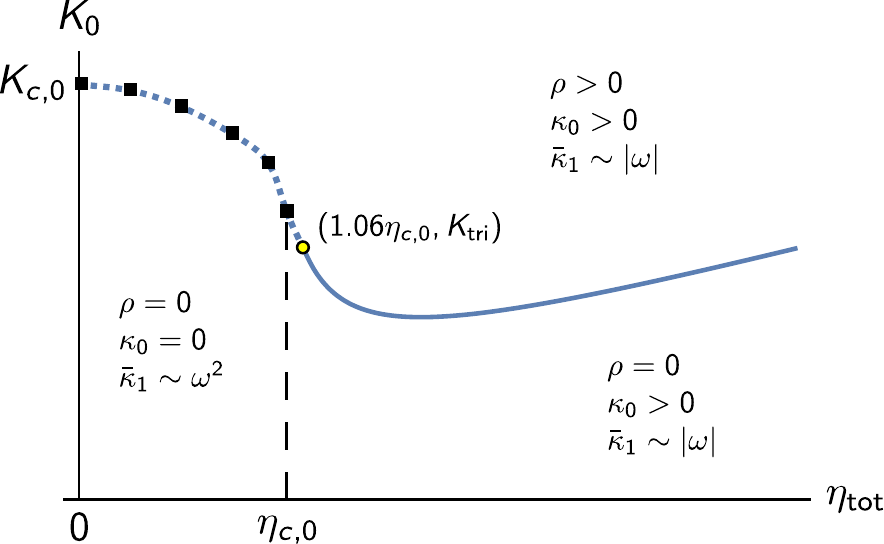}
    \caption{Mean-field ($N\to\infty$) phase diagram of the $(2+1)$-D quantum theory at zero temperature as a function of disorder strength $\etatot$ and CDW ordering parameter $K_0$ for $g_R/g_c = 2$ (in the $SU(N)$ symmetric phase for clean decoupled $\mathbb{C}P^N$ models at zero temperature). The three regimes discussed in the text are labeled by the two order parameters $\rho$ (CDW amplitude) and $\kappa_0$ (static response strength). $\bar{\kappa}_1(\omega)$ is the subtracted self-energy defined below Eq. \eqref{eq:kappa1Eq}. The black squares are first order CDW transitions found numerically from solutions of Eqs. \eqref{eq:dirtysaddleszerotemp} and the dashed curve joining them is a guide for the eye. The yellow circle is the tricritical point and the solid blue curve to the right of it is the exact phase boundary $K_c(\etatot)$ given by Eq. \eqref{eq:dirtycritKzero}; note that $K_{\mathrm{tri}} = K_c(\eta_{\mathrm{tri}})$. The vertical dashed black line is the weak-to-strong disorder crossover given by Eq. \eqref{eq:critdisorder}. $K_{c,0} = 8\pi\mu(1-g_c/g_R)$ is the CDW transition in the clean system.}
    \label{fig:phase_diagram_dirty}
\end{figure}

Finally, we consider the behavior of the system as a function of disorder in the regime $\rho > 0$. Solving Eqs. \eqref{eq:dirtysaddleszerotemp} numerically for $\rho$ and $\kappa_0$ yields the plots shown in Fig. \ref{fig:disorderdependence}. The behavior of the CDW amplitude $\rho$ at weak disorder is consistent with our previous observations of a first order transition at weak disorder, whereas when approaching the zero amplitude critical disorder $\eta_{c,0}(K_0)$, defined by inverting Eq. \eqref{eq:dirtycritKzero}, the transition is clearly continuous. The most interesting behavior, however, is that of the static response strength $\kappa_0$. On a logarithmic scale, it is clear that $\kappa_0$ never vanishes for any finite disorder, though it approaches zero as $\etatot \rightarrow 0$. This is in stark contrast with the behavior when $\rho = 0$, where we have shown there exists a critical disorder $\eta_{c,0}$ at which $\kappa_0$ vanishes. Note that we find this behavior across a range of values of $g_R$ and $K_0$. This implies that, unlike in the classical model, there is no weak disorder regime when $\rho > 0$ (c.f. Fig. 2 of Ref. \cite{OBrien2024}). While this may seem surprising, we can still draw intuition from the classical model. In that case, the critical disorder $\eta_{c}$ was suppressed continuously with increasing $K_0$, as it took less pinning from the disorder to disrupt the pairing of the $\mathbb{C}P^N$ fields forming the CDW condensate. In the present case, it is clear that the first order transition immediately overcomes the capacity of the system to remain ``clean'', forcing a simultaneous transition into the strong disorder regime. This behavior has parallels with the clean system discussed earlier in this work, where the $U(1)$ CDW transition immediately precipitates the $SU(N)$ transition; there is no intermediate vestigial phase. The results of this section are summarized in the mean-field phase diagram in Fig. \ref{fig:phase_diagram_dirty}.

\subsubsection{Order Parameter Correlations with Strong Disorder}

We now consider the leading order fluctuations of the order parameter about the $N=\infty$ mean-field state. Just as in Ref. \cite{OBrien2024}, we emphasize that mean-field properties of the weak disorder regime are largely artifacts of the large-$N$ limit which are cured by including $1/N$ corrections. Therefore, in this section we will focus on the more physically relevant regime with $\kappa_0 > 0$ and comment on the $\kappa_0 = 0$ case afterwards. When we also have $\rho = 0$, we use the Cartesian representation $\sigma(x) = \sigma^R(x) + i \sigma^I(x)$ and expand the effective action Eq. \eqref{eq:disorderedEffAct} to quadratic order to obtain a low energy effective theory
\begin{equation}
    S_{\mathrm{eff}} = \frac{N}{2} \sum_{a=R,I} \sum_{i,j=1}^n \int \frac{\dd^2 \vb{p}}{(2\pi)^2}  \frac{\dd \omega}{2\pi}  \, \sigma^{(a)}_i(-\vb{p},-\omega) \left[ \Pi^{(1)}_\sigma(\vb{p},\omega;n) \hat{I} - \Pi^{(2)}_\sigma(\vb{p},\omega;n) \hat{X} \right]_{ij} \sigma^{(a)}_j(\vb{p},\omega), \label{eq:OP_effAct_disordered}
\end{equation}
where the kernels are
\begin{subequations}
\begin{align}
    \Pi_\sigma^{(1)}(\vb{p},\omega;n) - \Pi_\sigma^{(2)}(\vb{p},\omega;n) &= \frac{2}{K_0} - 2\int \frac{\dd^2 \vb{p}'}{(2\pi)^2} \frac{\dd \omega'}{2\pi} G^{ii}(\vb{p}',\omega') G^{ii}(\vb{p} + \vb{p}',\omega + \omega'), \\
\begin{split}
    \Pi_\sigma^{(2)}(\vb{p},\omega;n) &= 2 \int \frac{\dd^2 \vb{p}'}{(2\pi)^2} \frac{\dd \omega'}{2\pi} G^{ij}(\vb{p}',\omega') G^{ij}(\vb{p} + \vb{p}',\omega + \omega') ,
\end{split}
\end{align}
\end{subequations}
and the $N=\infty$ propagator is
\begin{equation}
    \hat{G}^{-1}(\vb{p},\omega) =  [\omega^2 + \vb{p}^2 + M^2 + \bar{\kappa}_1(\omega)] \hat{I} - \kappa_0 (2\pi) \delta(\omega) \hat{X} ,
\end{equation}
where $\hat{I}$ is the $n\times n$ identity matrix and $\hat{X}$ is the $n\times n$ matrix with a one in every entry. The disorder-averaged propagator of the $\vb*{z}$ and $\vb*{w}$ fields in this regime is
\begin{equation}
    \overline{G(\vb{p},\omega)} = \lim_{n\rightarrow0} \frac{1}{n} \tr \hat{G}(\vb{p},\omega) = \frac{1}{\omega^2 + \vb{p}^2 + M^2 + \bar{\kappa}_1(\omega)} + \frac{\kappa_0 (2\pi) \delta(\omega)}{(\vb{p}^2 + M^2)^2},
\end{equation}
where the interpretation of $\kappa_0$ as the static response strength is made manifest as the coefficient of the frequency delta function. It follows that the disorder-averaged order parameter correlation function is
\begin{align}
    \overline{\langle \sigma(\vb{p},\omega) \sigma^*(-\vb{p},-\omega) \rangle} &= \lim_{n\rightarrow0} \frac{1}{n} \tr\langle \sigma_i(\vb{p},\omega) \sigma_j^*(-\vb{p},-\omega) \rangle \nn \\
    &= 2\left(\frac{1}{\Pi^{(1)}_\sigma(\vb{p},\omega;0)} + \frac{\Pi^{(2)}_\sigma(\vb{p},\omega;0)}{[\Pi^{(1)}_\sigma(\vb{p},\omega;0)]^2} \right). \label{eq:disorderaveragedprop}
\end{align}
First, it is simple to obtain
\begin{equation}
    \Pi_\sigma^{(2)}(0,\omega;0) = \frac{\kappa_0^2}{6\pi M^6} (2\pi) \delta(\omega) = \frac{32\pi^2 \kappa_0^2}{3 \eta_{\mathrm{tot}}^6} (2\pi) \delta(\omega),
\end{equation}
from which we identify that the second term in Eq. \eqref{eq:disorderaveragedprop} is the ubiquitous static double-Lorentzian contribution \cite{Imry1975,Pytte1981,Yoshizawa1982}. This behavior is identical to that found in the quantum NLSM with quenched disorder; see Appendix \ref{app:nlsm}. Next, we can divide the integral defining $\Pi_\sigma^{(1)}(\vb{p},\omega;n)$ into two terms: (a) quasiparticle scattering which is purely due to the replica identity term in the propagator and (b) impurity scattering which has a contribution from the static all-to-all replica term,
\begin{subequations}
\begin{align}
\begin{split}
    \Pi_\sigma^{(1,a)}(\vb{p},\omega;0) &= 2\int \frac{\dd^2 \vb{p}'}{(2\pi)^2} \frac{\dd \omega'}{2\pi}  \frac{1}{\omega^{\prime 2} + \vb{p}^{\prime 2}+M^2 + \bar{\kappa}_1(\omega')} \\
    &\kern8em \times\frac{1}{(\omega + \omega')^2 + (\vb{p} + \vb{p}')^2 + M^2 + \bar{\kappa}_1(\omega + \omega')}, 
\end{split}\\
    \Pi_\sigma^{(1,b)}(\vb{p},\omega;0) &= 4 \kappa_0 \int \frac{\dd^2 \vb{p}'}{(2\pi)^2} \frac{1}{(\vb{p}^{\prime 2} + M^2)^2} \frac{1}{\omega^2 + (\vb{p} + \vb{p}')^2 + M^2 + \bar{\kappa}_1(\omega)} .
\end{align}
\end{subequations}
Keeping contributions up to $\mathcal{O}(\omega^2)$ in the self-energy function $\bar{\kappa}_1(\omega)$, we find that
\begin{subequations}
\begin{align}
    \Pi_\sigma^{(1,a)}(\vb{p},\omega;0) &\simeq \frac{1}{\sqrt{6\pi} \eta_{\mathrm{tot}}} - \frac{\sqrt{2}(\sqrt{12}\pi - 9)}{9 \sqrt{\pi} \eta_{\mathrm{tot}}^3} \vb{p}^2 - \frac{\sqrt{2} \left(\sqrt{48} \pi -9\right)}{27 \sqrt{\pi} \eta_{\mathrm{tot}} ^3} \omega^2,\\
    \Pi_\sigma^{(1,b)}(\vb{p},\omega;0) &\simeq \frac{8\pi \kappa_0}{\eta_{\mathrm{tot}}^4} - \frac{32 \pi^2 \kappa_0}{3 \eta_{\mathrm{tot}}^6} \vb{p}^2  - \frac{32\sqrt{2 \pi^3} \kappa_0}{3 \eta_{\mathrm{tot}}^5} \abs{\omega} + \frac{160 \pi^2 \kappa_0}{9 \eta_{\mathrm{tot}}^6} \omega^2 .
\end{align}
\end{subequations}
Therefore, at least at zero temperature, the dynamic quasiparticle scattering is analytic in frequency despite the damping of $\vb*{z}$ and $\vb*{w}$ excitations, while the impurity scattering displays damping. Finally, we can put this all together by analytically continuing to real frequency (using the definition of the absolute value $\abs{z} = \sqrt{z^2}$). Then, the disorder-averaged dynamic susceptibility of the order parameter takes the form
\begin{equation}
    \overline{\langle \sigma(\vb{p},-i\omega) \sigma^*(-\vb{p},i\omega) \rangle} \simeq \frac{1}{m_\sigma^2 + \gamma_\sigma (\vb{p}^2 - v_\sigma^{-2} \omega^2) - i \Gamma_\sigma \abs{\omega}} + \frac{ \eta^2_\sigma (2\pi) \delta(\omega)}{(m_\sigma^2 + \gamma_\sigma \vb{p}^2)^2},
\end{equation}
and the parameters are found to be
\begin{align}
\begin{split}
    m_\sigma^2 = \frac{1}{K_0} - \frac{1}{2\sqrt{6\pi} \eta_{\mathrm{tot}}} - \frac{4\pi \kappa_0}{\eta_{\mathrm{tot}}^4}, \kern 3em \gamma_\sigma = \frac{(\sqrt{12}\pi - 9)}{9 \sqrt{2\pi} \eta_{\mathrm{tot}}^3} + \frac{16 \pi^2 \kappa_0}{3 \eta_{\mathrm{tot}}^6}, \\
    \gamma_\sigma v_\sigma^{-2} = \frac{\left(\sqrt{48} \pi -9\right)}{27 \sqrt{2\pi} \eta_{\mathrm{tot}} ^3} - \frac{80 \pi^2 \kappa_0}{9 \eta_{\mathrm{tot}}^6}, \kern3em \Gamma_\sigma = \frac{16\sqrt{2 \pi^3} \kappa_0}{3 \eta_{\mathrm{tot}}^5}, \kern3em \eta_\sigma^2 = \frac{16\pi^2 \kappa_0^2}{3 \etatot^6}.
\end{split}
\end{align}
Given $\kappa_0(\etatot)$ in Eq. \eqref{eq:disorderamplitude1}, we can identify a frequency scale $\omega_{\mathrm{damp}}$ above which the order parameter correlations are under-damped; for $\etatot$ close to $\eta_{c,0}$, $\omega_{\mathrm{damp}} \sim (g_R^{-1} - g_{c}^{-1})^2(\etatot - \eta_{c,0})/\etatot^2$, while for $\etatot \gg \eta_{c,0}$, $\omega_{\mathrm{damp}} \sim \etatot$. In general, however, the low-frequency behavior is always over-damped. Note that the mass gap $m_\sigma$ derived here comes from a low-frequency expansion of $\bar{\kappa}_1(\omega)$, whereas the exact expression is simply $m_\sigma^2 = K_0^{-1} - K_c^{-1}(\etatot)$, where $K_c(\etatot)$ is given by Eq. \eqref{eq:dirtycritKzero}. Finally, we observe that since $\Gamma_\sigma \propto \kappa_0$, the damping vanishes as the disorder approaches the critical value $\eta_{c,0}$ and the static response strength vanishes.

When there is a finite CDW condensate $\rho > 0$, we can write down an effective action for the phase $\theta(x)$ of the CDW order parameter $\sigma(x) = \rho(x) e^{i\theta(x)}$,
\begin{equation}
    S_{\mathrm{eff}} = \frac{N}{2} \sum_{i,j=1}^n \int \frac{\dd^2 \vb{p}}{(2\pi)^2}  \frac{\dd \omega}{2\pi}  \, \theta_i(-\vb{p},-\omega) \left[ \Pi^{(1)}_\theta(\vb{p},\omega;n) \hat{I} - \Pi^{(2)}_\theta(\vb{p},\omega;n) \hat{X} \right]_{ij} \theta_j(\vb{p},\omega). \label{eq:OP_effAct_disordered_phase}
\end{equation}
Even in the classical theory, the nature of the regime with $\rho > 0$ is extremely complex \cite{Houghton1981,Goldschmidt1982}, so a full treatment of this problem is beyond the scope of this present work and will be left to a future publication. However, we wish to highlight the following interesting observation: Since the effective action Eq. \eqref{eq:disorderedEffAct} has an average $U(1)$ symmetry corresponding to the replica-diagonal subgroup of the replicated $U(1)^n$ symmetry group, a consistent low energy effective theory must satisfy the associated Ward identity. This implies that the kernel in Eq. \eqref{eq:OP_effAct_disordered_phase} must have an eigenvalue which vanishes at $\abs{\vb{p}} = \omega = 0$. Together with a non-analytic self-energy which we have shown produces damping, the existence of a gapless mode is suggestive of diffusive quasiparticle transport.

\subsubsection{Correlations in the Weak Disorder Regime}

\begin{figure}[!t]
    \centering
    \includegraphics[width=0.9\linewidth]{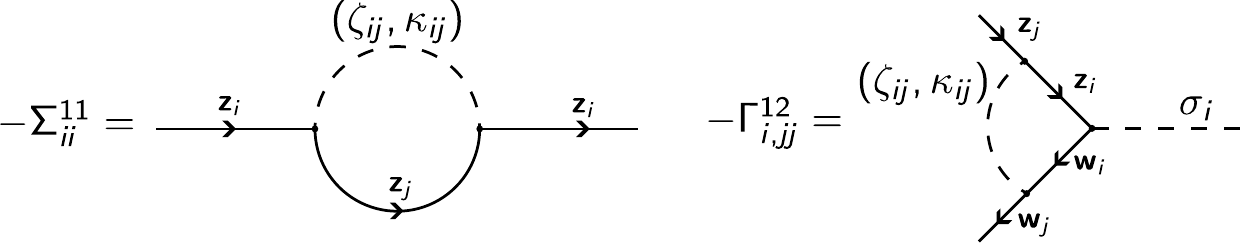}
    \caption{One-loop self energy and three-point vertex contributions from fluctuations of the Hubbard-Stratonovich fields $\zeta_{ij}$ and $\kappa_{ij}$ which decouple the quartic interactions induced by averaging over disorder configurations.}
    \label{fig:vertexcorrection}
\end{figure}

In the weak disorder regime with $\kappa_0 = 0$, the $N=\infty$ propagator is diagonal in the replica index and does not contain a static peak
\begin{equation}
    \hat{G}^{-1}(\vb{p},\omega) =  [\omega^2 + \vb{p}^2 + M^2 + \bar{\kappa}_1(\omega)] \hat{I} .
\end{equation}
Therefore, the leading contribution to the low energy effective theory is
\begin{equation}
    S_{\mathrm{eff}} = \frac{N}{2} \sum_{a=R,I} \sum_{i=1}^n \int \frac{\dd^2 \vb{p}}{(2\pi)^2}  \frac{\dd \omega}{2\pi}  \, \sigma^{(a)}_i(-\vb{p},-\omega)  \Pi_\sigma(\vb{p},\omega) \sigma^{(a)}_i(\vb{p},\omega), \label{eq:OP_effAct_disordered_weak}
\end{equation}
where the kernel is
\begin{equation}
    \Pi_\sigma(\vb{p},\omega) = \frac{2}{K_0} - 2\int \frac{\dd^2 \vb{p}'}{(2\pi)^2} \frac{\dd \omega'}{2\pi} G^{ii}(\vb{p}',\omega') G^{ii}(\vb{p} + \vb{p}',\omega + \omega').
\end{equation}
This implies that the disorder-averaged dynamic susceptibility of the order parameter is
\begin{align}
    \overline{\langle \sigma(\vb{p},-i\omega) \sigma^*(-\vb{p},i\omega) \rangle} = \frac{2}{\Pi_\sigma(\vb{p},-i\omega)} \simeq \frac{1}{m_\sigma^2 + \gamma_\sigma(\vb{p}^2 - v_\sigma^{-2}\omega^2)},\label{eq:disorderaveragedpropweak}
\end{align}
where the parameters are
\begin{equation}
    m_\sigma^2 \simeq \frac{1}{K_0} - \frac{1}{8\pi M}\sqrt{1 - \frac{\etatot^2}{4\pi M^2}}, \kern2em \gamma_\sigma \simeq \frac{1}{96\pi M^3}\sqrt{1 - \frac{\etatot^2}{4\pi M^2}}, \kern2em v_\sigma^2 \simeq 1 - \frac{\etatot^2}{4\pi M^2} ,
\end{equation}
and $M$ here is a function of $\etatot$ determined from the solution of the saddle-point equation Eq. \eqref{eq:gappsaddle}; see Fig. \ref{fig:cpn_gap} in Appendix  \ref{app:cpn}. Note that we only kept the low frequency contribution of the self-energy $\bar{\kappa}_1(\omega)$, so these expressions are not valid when $\etatot$ is too close to $\eta_{c,0}$. Unlike in the classical model discussed in Ref. \cite{OBrien2024}, the disorder strength $\etatot$ \textit{does} enter into observables in the weak disorder regime through the self-energy $\bar{\kappa}_1(\omega)$. However, the lack of a double-Lorentzian term in Eq. \eqref{eq:disorderaveragedpropweak} implies a weak infrared singularity in momentum corresponding to a lower critical dimension $d=2$ which appears to violate the Imry-Ma theorem \cite{Imry1975}. As explained in Ref. \cite{OBrien2024}, this apparent violation is an artifact of the $N\to\infty$ limit which is cured by including vertex corrections at order $1/N$: In general, the replica off-diagonal kernel for the order parameter $\Pi_{\sigma,ij}^{(2)}(p)$ has the form
\begin{equation}
    -\Pi_{\sigma,ij}^{(2)}(p) = \int \frac{\dd^3 q}{(2\pi)^3} \Gamma^{ab}_{i,k_1k_2}(p,-k,q) \Gamma^{cd}_{j,\ell_1\ell_2}(-p,k,-q) G^{ac}_{k_1\ell_1}(k) G^{bd}_{k_2\ell_2}(q) (2\pi)^3 \delta^{(3)}(p+q-k) \label{eq:kernelgeneralform}
\end{equation}
where $a,b,c,d=1,2$ denote either $\vb*{z}$ or $\vb*{w}$, and $i,k_1,k_2,\ell_1,\ell_2$ are replica indices, with implied summation over repeated indices. Importantly, $G^{ab}_{ij}(p)$ is the exact propagator for the $SU(N)$ $\vb*{z}$ and $\vb*{w}$ fields and $\Gamma^{ab}_{i,jk}(p_1,p_2,p_3)$ is the exact three-point vertex between $\sigma_i$ and the $\vb*{z}_i$ and $\vb*{w}_j$ fields. In the large-$N$ technique, these exact functions have self-consistent perturbative expansions in $1/N$, so that even though $\Pi_{\sigma,ij}^{(2)}(p) = 0$ as $N\to\infty$ in the weak disorder regime, higher-order corrections such as those represented by the Feynman diagrams in Fig. \ref{fig:vertexcorrection} will generate replica off-diagonal interactions, leading to a double-Lorentzian peak.

However, the nature of the $1/N$ expansion also implies that in the quantum system the weakly- and strongly-disordered regimes remain fundamentally distinct, unlike in the classical case. Since the vertex corrections involve integrals of products of propagators, the absence of a static delta function term in the $N=\infty$ propagator at weak disorder implies that the double-Lorentzian contribution at order $1/N$ will not be static. While vertex corrections in the strong disorder regime may produce dynamic contributions to the double-Lorentzian term, it is the existence of the static delta function in this case which distinguishes the two regimes. Secondly, due to the absence of non-analytic $\abs{\omega}$ behavior in the weak disorder self-energy as $N\to\infty$, if damping is to emerge at order $1/N$ it must be due to more complex interactions such as multi-particle branch cuts.

\subsection{Finite Temperature \label{sec:finitetempdisorder}}

\subsubsection{Large-\texorpdfstring{$N$}{N} Solution}

We now consider the disordered system at finite temperature. The saddle-point equations in Eqs. \eqref{eq:disorderedsaddles} are unchanged apart from the usual replacement
\begin{equation}
    \int_{-\infty}^\infty \frac{\dd \omega}{2\pi} \longrightarrow \frac{1}{\beta} \sum_{\omega_n},
\end{equation}
where $\omega_n = 2\pi n/\beta$ are the Matsubara frequencies. Notably, because the equations for the bilocal fields do not involve any frequency integrals, the solutions for $\tilde{\kappa}_0(\omega)$ and $\bar{\kappa}_1(\omega)$ found above are unchanged by temperature. Before discussing the impact of thermal fluctuations on the order parameter correlations, we can derive how the mean-field phase boundaries we calculated above are shifted as a function of temperature. 

At fixed $g_R$ and $\eta_{\mathrm{tot}}$, for $K_0<K_c$, that is, when the CDW amplitude $\rho = 0$, the saddle-point Eq. \eqref{eq:saddleg0} implies
\begin{equation}
    \int \frac{\dd^d \vb{p}}{(2\pi)^d} \left[ \int\frac{\dd \omega}{2\pi}   \frac{1}{\omega^2 + \vb{p}^2 + M_0^2 + \bar{\kappa}_1(\omega) } - T \sum_{\omega_n} \frac{1}{\omega_n^2 + \vb{p}^2 + M_T^2 + \bar{\kappa}_1(\omega_n)} \right]  = \frac{\kappa_T - \kappa_0}{\eta_{\mathrm{tot}}^2} ,\label{eq:saddlefinitetemp}
\end{equation}
where, in general, the thermal gap $M_T$ depends on temperature, and $M_0$ and $\kappa_0$ are the zero temperature gap and static response strength. However, in the strong disorder regime $M^2 = \etatot^2/4\pi$ independently of temperature. Therefore, after applying the Poisson summation formula,
\begin{equation}
    \kappa_T - \kappa_0 = -2 \etatot^2 \sum_{k=1}^\infty \int \frac{\dd^d \vb{p}}{(2\pi)^d} \int \frac{\dd z}{2\pi} \frac{e^{-ik\beta z}}{z^2 + \vb{p}^2 + M^2 + \bar{\kappa}_1(z)}.
\end{equation}
This expression allows us to identify the low temperature behavior with the small $z$, that is, large $k\beta$ behavior of the Fourier transform, and we find
\begin{equation} \label{eq:disorderdensitytemp}
    \kappa_T - \kappa_0 \simeq - \sqrt{\frac{\pi}{18}}\eta_{\mathrm{tot}} T^2 ,
\end{equation}
which implies a critical temperature
\begin{equation} \label{eq:disordercrittemp}
    T_c \simeq \left(\frac{18}{\pi} \right)^{1/4} \sqrt{\etatot \left(\frac{1}{g_R} - \frac{1}{g_c} \right) + \mathsf{c}_1 \etatot^2 },
\end{equation}
above which the system is no longer in the strongly-disordered regime. This is physically intuitive, as strong thermal fluctuations will overcome the ``pinning'' tendency of the quenched disorder. This result is also interesting because it highlights the important effect of the linear in frequency self-energy $\bar{\kappa}_1(\omega)$; ignoring the damping would yield a drastically different temperature dependence $\kappa_T - \kappa_0 \sim - \etatot^2 T e^{-\etatot^2/4\pi T}$; damping enhances the effect of the thermal fluctuations. We can also investigate the high temperature limit of Eq. \eqref{eq:saddlefinitetemp}, which yields
\begin{equation}
    \frac{\kappa_T}{\eta_{\mathrm{tot}}^2 T} \simeq \frac{1}{g_R T} + \frac{1}{4\pi} \ln(\frac{\eta_{\mathrm{tot}}^2}{4\pi \mu^2}) ,
\end{equation}
which, up to re-definitions of coupling constants to absorb factors of $T$, is exactly the functional dependence of $\kappa_0$ in the classical $(2+0)$D theory \cite{OBrien2024}.

Similarly, the phase boundary for the formation of the CDW amplitude at fixed $g_R$ and $\etatot$ takes the form
\begin{align}
    \frac{1}{K_c(T)} - \frac{1}{K_c(0)}  &= \frac{\kappa_T - \kappa_0}{4\pi \eta_{\mathrm{tot}}^4} + \frac{1}{4\pi} \left[ \frac{1}{\beta} \sum_{\omega_n} \frac{1}{\omega_n^2 + M^2 + \bar{\kappa}_1(\omega)} - \int \frac{\dd \omega}{2\pi} \frac{1}{\omega^2 + M^2 + \bar{\kappa}_1(\omega)} \right] \nn \\
    & \simeq \left(\frac{2\sqrt{2 \pi^3}}{3} - \frac{1}{12 \sqrt{2\pi}} \right) \frac{T^2}{\etatot^3} - \frac{32 \sqrt{2\pi^9}}{45} \frac{T^4}{\etatot^5}. 
\end{align}
Therefore, at the lowest temperatures, the competition between thermal fluctuations and disorder actually decrease $K_c$, though this trend reverses at $T/\etatot \approx 0.12$, above which the critical coupling increases with temperature. As we have seen several times, for example, in Figs. \ref{fig:phase_diagram_clean2} and \ref{fig:phase_diagram_dirty}, reentrant order appears to be a feature of this theory.

Finally, we note that the high temperature limit of the phase diagram must be consistent with the $(2+0)$D classical theory studied in Ref. \cite{OBrien2024}. In particular, the weak disorder regime with a finite CDW amplitude ($\rho >0$ and $\kappa_0 = 0$) which we showed in the previous section was absent at $T=0$ must emerge as the temperature is increased. This is entirely consistent with the existence of the critical temperature Eq. \eqref{eq:disordercrittemp}; unfortunately an analogous analytic expression for the critical temperature in the regime where $\rho > 0$ does not exist due to the complicated temperature dependence of $\rho$.

\subsubsection{Order Parameter Correlations}

We now return to the dynamics of the order parameter and include the effects of thermal fluctuations. Compared to the zero temperature case, only the calculation of the dynamic quasiparticle scattering term $\Pi_\sigma^{(1,a)}$ is affected by the change to Matsubara summation, since none of the other contributions involved an integral over frequency. First, we use the Poisson summation formula to write
\begin{align} \label{eq:quasiparticlekernelmatsubara}
\begin{split}
    \Pi_\sigma^{(1,a)}(\vb{p},\omega_n;T,0) &= 2 \sum_{k=-\infty}^\infty \int \frac{\dd z}{2\pi} \frac{\dd^2 \vb{p}'}{(2\pi)^2} \frac{1}{z^{2} + \vb{p}^{\prime 2}+M^2 + \bar{\kappa}_1(z)} \\
    &\kern4em \times \frac{1}{(\omega_n + z)^2 + (\vb{p} + \vb{p}')^2 + M^2 + \bar{\kappa}_1(\omega_n + z)} e^{-i k \beta z} ,
\end{split}
\end{align}
where the $k=0$ term in the sum is precisely the $T=0$ contribution we have already evaluated in a previous section. To extract the low frequency response, we only need the leading behavior of the self-energy $\bar{\kappa}_1(z) \propto \abs{z}$. As is usual, the external (in the Feynman diagram loop sense) imaginary frequency $\omega_n$ is taken to be arbitrary and not discrete, so that $\bar{\kappa}_1(\omega_n + z) \propto \abs{\omega_n + z}$ is a sensible approximation. After analytic continuation, to leading order in real frequency and $1/\beta$, we find the thermal contribution
\begin{equation}
\begin{split}
    &\Pi_\sigma^{(1,a)}(\vb{p},-i\omega;T,0) - \Pi_\sigma^{(1,a)}(\vb{p},-i\omega;0,0) \simeq \frac{4 \sqrt{2} \pi ^{3/2}}{3 \beta ^2 \eta_{\mathrm{tot}} ^3} - \frac{16 \sqrt{2} \pi ^{5/2} }{9 \eta_{\mathrm{tot}} ^5 \beta ^2} \vb{p}^2  -\frac{2 \sqrt{2 \pi }}{ \eta_{\mathrm{tot}} ^3 \beta} i \abs{\omega} .
\end{split}
\end{equation}
We neglect the $\mathcal{O}(\omega^2)$ contribution since the low frequency behavior is always over-damped. Adding this to the earlier results, we have the disorder-averaged dynamic susceptibility of the order parameter
\begin{equation}
    \overline{\langle \sigma(\vb{p},-i\omega) \sigma^*(-\vb{p},i\omega) \rangle} \simeq \frac{1}{m_\sigma^2 + \gamma_\sigma\vb{p}^2 - i \Gamma_\sigma \abs{\omega}} + \frac{ \eta^2_\sigma (2\pi) \delta(\omega)}{(m_\sigma^2 + \gamma_\sigma \vb{p}^2)^2},
\end{equation}
where the parameters are
\begin{align}
\begin{split}
    m_\sigma^2 = \frac{1}{K_0} - \frac{1}{2\sqrt{6\pi} \eta_{\mathrm{tot}}} - \frac{4\pi \kappa_T}{\eta_{\mathrm{tot}}^4} - \frac{2 \sqrt{2} \pi ^{3/2} T^2}{3 \eta_{\mathrm{tot}} ^3}, \kern 3em  \Gamma_\sigma = \frac{16\sqrt{2 \pi^3} \kappa_T}{3 \eta_{\mathrm{tot}}^5} + \frac{\sqrt{2 \pi } T}{ \eta_{\mathrm{tot}} ^3} ,\\
    \gamma_\sigma = \frac{(\sqrt{12}\pi - 9)}{9 \sqrt{2\pi} \eta_{\mathrm{tot}}^3} + \frac{16 \pi^2 \kappa_T}{3 \eta_{\mathrm{tot}}^6} + \frac{8 \sqrt{2} \pi ^{5/2} T^2}{9 \eta_{\mathrm{tot}} ^5}, \kern3em \eta_\sigma^2 = \frac{16\pi^2 \kappa_T^2}{3 \etatot^6},
\end{split}
\end{align}
where the finite temperature static response strength $\kappa_T$ is given by Eq. \eqref{eq:disorderdensitytemp}. With the additional dynamic quasiparticle scattering contribution at finite temperature, the damping factor $\Gamma_\sigma$ no longer vanishes smoothly in the limit that $\etatot$ approaches the critical value $\eta_{c,0}(T)$ (invert Eq. \eqref{eq:disordercrittemp} for $\etatot$ as a function of $T$). This reflects the discontinuity in the self-energy $\bar{\kappa}_1(\omega)$ at the critical disorder which is evident in the low frequency expansion in Eq. \eqref{eq:self-energy}. Therefore, for $T>0$ there exists a finite frequency scale $\omega_{\mathrm{damp}} \sim T$ below which the order parameter correlations are over-damped. This is a significant difference compared to the quantum NLSM at finite temperature discussed in Appendix \ref{app:nlsm}, for which the $N=\infty$ order parameter correlation function is completely undamped; we note that damping can still manifest in the NLSM at higher order in $1/N$.

\section{Discussion \label{sec:disc}}

In this paper, we have extended the model we introduced in Ref. \cite{OBrien2024} to explore the interplay of quantum and thermal fluctuations with quenched random disorder in two-dimensional incommensurate charge density waves. We derived the structure of the phase diagram non-perturbatively as a function of temperature and disorder strength in the large-$N$ limit, as well as the dynamics of the CDW order parameter fluctuations. Importantly, by representing the CDW as a composite $U(1)$ order parameter of the parent $SU(N)$ fields of our model, we ensured that the clean limit of the theory remains consistent; previous large-$N$ studies directly generalized the $U(1)$ CDW order parameter to $U(N)$, and hence, did not capture the fact that at finite temperature in the absence of disorder a $U(1)$ order parameter should have a BKT transition \cite{Nie2014}. Additionally, our approach is purposefully general, and can be used to describe any system with a $U(1)$ order parameter coupled to random field disorder. We note that throughout this work, we have only considered the regime in which replica permutation symmetry is unbroken. We showed in Ref. \cite{OBrien2024}, that our model does allow a range of parameters for which the replica-symmetric ground state is unstable. However, the glassy physics associated with replica symmetry breaking (e.g., Griffiths singularities due to rare configurations of disorder \cite{Griffiths1969}, or multifractal probability distribution for observables \cite{Wiese2022}) cannot naturally be incorporated into the $1/N$ expansion, and, as such, lies outside the scope of the present work.

Our first main result is the large-$N$ phase diagram of the quantum theory as a function of temperature and coupling between the $SU(N)$ fields in the absence of disorder. At zero temperature we found that the CDW order always coexists with the parent $SU(N)$ order and that the CDW amplitude has a first order discontinuity at the phase transition. This behavior stood in contrast with the continuous transition we described in the two-dimensional classical model in Ref. \cite{OBrien2024}. However, after turning on a finite temperature, we showed that in the large-$N$ limit, thermal fluctuations soften the first order jump so that beyond a tricritical point the CDW amplitude forms continuously. To make contact with the classical limit, we noted that the actual phase transition in this regime will lie in the 2D BKT universality class (with parameters renormalized by quantum fluctuations). We also computed the correlation function of the CDW order parameter and drew parallels with the large-$N$ solution of the quantum nonlinear sigma model \cite{Chubukov1994}, specifically the lack of damping to leading order in $1/N$. 

We then considered the primary focus of this work: quantum dynamics of the CDW in the presence of quenched random field disorder. We showed how averaging over disorder configurations using the replica trick produces interactions which are nonlocal in time. Solving the disordered theory in the large-$N$ limit, we found that the local part of the interaction acted as in the classical theory by inducing a crossover from weakly- to strongly-disordered behavior. Our second main result is that the new nonlocal part of the interaction causes the disorder crossover to also manifest as a change in the dynamics of the order parameter, with under-damped dynamics when the disorder is weak and over-damped dynamics when the disorder is strong. We also showed that, in the large-$N$ limit, at zero temperature the damping is only due to scattering from the static disorder. The drastic change in behavior of the dynamics should be observable in measurements of the dynamic structure factor or susceptibility of the CDW order parameter as a strongly disorder dependent broadening of the line-shape. While we anticipate that sub-leading $1/N$ corrections to our model may also allow for damping when the disorder is weak, those would be higher-order effects which we do not expect to significantly affect the evolution of experimental line-shapes through the crossover. We also argued on the basis of symmetry that the Goldstone mode which exists in the regime of the clean theory with a well-formed amplitude of the CDW order parameter might become diffusive in the presence of disorder. However, it is well-known that at finite temperature this regime exhibits a complicated competition between disorder and vortices \cite{Houghton1981,Goldschmidt1982}, so questions of diffusive transport are left for a future work.

Finally, we used our model to understand the interplay of quantum and thermal fluctuations in the presence of disorder. After mapping out the complete large-$N$ phase diagram as a function of the disorder strength and the coupling between the $SU(N)$ fields, we calculated the temperature dependence of phase boundaries in the low temperature limit. This complements our results in Ref. \cite{OBrien2024}, which correspond to the high-temperature limit where quantum effects are completely suppressed. We also showed that scattering processes which are undamped (in the large-$N$ limit) both at finite temperature in the absence of disorder and at zero temperature with disorder acquire a dissipative contribution at finite temperature in the presence of disorder. In this way, our theory demonstrates that thermal fluctuations enhance the disorder-induced damping of the CDW order parameter.

In this work, we have considered charge density waves without any coupling to underlying electronic degrees of freedom. This is an appropriate approximation when electron-electron interactions are strong enough to form an insulating CDW state and when all energy scales are much less than the electronic gap. A complete quantum theory of charge density waves necessarily requires coupling the CDW to a Fermi surface, as was done in the absence of disorder by one of us in Ref. \cite{Sun2008}. This is a highly non-trivial problem, given the need to account for the effect of disorder on gapless charge carriers, ideally non-perturbatively. In particular, the fermions will generally lead to Landau damping of the CDW order parameter, and the interplay of this additional source of damping with the effects described in this work will lead to richer structure in the correlations of CDW fluctuations. Solving this open problem would help address many questions raised by recent experiments. 

Finally, to the best of our knowledge, this paper is also the first work to apply the large-$N$ technique to examine the physics of composite order parameters coupled to random disorder. We believe our results have implications for understanding the behavior of disordered composite and vestigial orders such as those found in models of pair density wave superconductors \cite{Berg2009} and, more broadly, to other phases of matter described by composite order parameters such as electronic nematic order \cite{Fradkin2010,Fernandes-2019}.

\begin{acknowledgments}
M C O thanks J Gliozzi and D Manning-Coe for useful discussions. This work was supported in part by the US National Science Foundation through the grant DMR 2225920 at the University of Illinois.
\end{acknowledgments}

\renewcommand{\theequation}{\Alph{section}.\arabic{equation}}

\appendix

\addcontentsline{toc}{section}{Appendix \ref*{app:nlsm}: Review of Quantum Nonlinear Sigma Model with Quenched Disorder}
\tocless{\section}{Review of Quantum Nonlinear Sigma Model with Quenched Disorder\label{app:nlsm}}

In this appendix, we review the large-$N$ solution of the quantum nonlinear sigma model (NLSM) with quenched random disorder at finite temperature. The action for the quantum NLSM is
\begin{equation}
    S[\vb*{h}] = \int \dd^d \vb{x} \dd\tau \left[ \frac{1}{2g} (\del_\mu \vb*{n}(\vb{x},\tau))^2 - \vb*{h}(\vb{x}) \cdot \vb*{n}(\vb{x},\tau)\right],\kern3em \vb*{n}^2(\vb{x},\tau) = 1,
\end{equation}
where $\vb*{h}(\vb{x})$ is a static external source. If $\vb*{h}(\vb{x}) = \frakh(\vb{x})$ is a random field distributed according to
\begin{equation}
    \overline{\mathfrak{h}^a(\vb{x})} = 0, \kern3em \overline{\mathfrak{h}^a(\vb{x}) \mathfrak{h}^b(\vb{y})} = \eta^2 \delta_{ab} \delta^{(d)}(\vb{x} - \vb{y}),
\end{equation}
then disorder averages can be calculated using the replica trick,
\begin{align}
    \overline{\mathcal{Z}^n} &= \int \mathcal{D} \mathfrak{h}\, \exp\left(- \int \dd^d \vb{x} \frac{\mathfrak{h}^2}{2\eta^2} \right) \mathcal{Z}[\mathfrak{h}]^n \nn \\
\begin{split}
    &= \int \prod_{j=1}^n \mathcal{D} \vb*{n}_j \mathcal{D} \lambda_j \, \exp\left(- \sum_{i,j=1}^n \int \dd^d \vb{x} \dd\tau \frac{1}{2g} \Big[ (\del_\mu \vb*{n}_j)^2 + \lambda_j (\vb*{n}^2 - 1) \Big] \delta_{ij}  \right. \\
    &\kern12em+ \left. \frac{\eta^2}{2} \sum_{i,j=1}^n \int \dd^d \vb{x} \dd \tau_1 \dd \tau_2\, \vb*{n}_i(\tau_1) \cdot \vb*{n}_j(\tau_2) \right) , \label{eq:disavgON}
\end{split}
\end{align}
where $\lambda_j$ are the Lagrange multipliers for the unit vector constraints $\vb*{n}_j^2 = 1$. Note that the disorder only couples the imaginary time averages of the order parameter with different replica indices. After integrating out the $\vb*{n}_j$ fields, we obtain the effective action
\begin{equation}
    S_{\mathrm{eff}}/N = \frac{1}{2} \Tr \ln\left(\delta(\tau_1 - \tau_2) [-\del^2 \hat{I} + \mathrm{diag}_r(\lambda)] - g_0\eta_0^2 \hat{X} \right) - \sum_{j=1}^n \int \dd^d \vb{x} \dd \tau \frac{\lambda_j}{2g_0} ,
\end{equation}
where $\mathrm{diag}_r(\lambda)$ is the matrix with the $\lambda_j$ along its diagonal entries, $\hat{I}$ is the $n\times n$ identity matrix, $\hat{X}$ is the $n\times n$ matrix with a one in every entry, $g = g_0/N$, and $g \eta^2 = g_0 \eta_0^2$. In the limit $N\rightarrow\infty$, the values of the $\lambda_j$ are determined by the saddle point equations. Assuming a replica-symmetric solution $\lambda_j = m^2$, we find in the replica limit $n\rightarrow 0$
\begin{equation}
    \frac{1}{\beta} \sum_{\omega_n} \int \frac{\dd^d \vb{q}}{(2 \pi)^d} \frac{1}{\omega_n^2 + \vb{q}^2 + m^2} + \int \frac{\dd^d \vb{q}}{(2 \pi)^d} \frac{g_0 \eta_0^2}{(\vb{q}^2 + m^2)^2}  = \frac{1}{g_0} .
\end{equation}
This implies that the disorder-averaged imaginary frequency propagator for the order parameter is
\begin{equation}
    \overline{G(\vb{p},\omega_n)} = \frac{1}{\omega_n^2 + \vb{p}^2 + m^2} + \frac{g_0 \eta_0^2}{(\vb{p}^2 + m^2)^2} \beta \delta_{n,0}.
\end{equation}
The two key features of this expression are: (i) The disorder contributes an Imry-Ma type double-Lorentzian term \cite{Imry1975}, which is purely static. (ii) The dynamics of the order parameter are completely unaffected by disorder as $N\to\infty$, and, notably, are underdamped, since $[\overline{G(\vb{p},-i\omega)}]^{-1} \propto \omega^2$.

\addcontentsline{toc}{section}{Appendix \ref*{app:cpn}: Quantum \texorpdfstring{$\mathbb{C}P^N$}{CPN} Model with Quenched Disorder}
\tocless{\section}{Quantum \texorpdfstring{$\mathbb{C}P^N$}{CPN} Model with Quenched Disorder\label{app:cpn}}

In the main body of this work, we considered a two-component generalization of the $\mathbb{C}P^N$ model. In this appendix, we present the solution of the simpler model in a quenched random field.

As explained in Ref. \cite{OBrien2024}, the minimal way to couple a $\mathbb{C}P^N$ model to quenched disorder is in the form of a random field $\mathfrak{z}^a(\vb{x})$ in the adjoint (vector) representation of $SU(N)$. Therefore, the partition function for a particular realization of disorder is
\begin{equation}
\begin{split}
    \mathZ[\mathfrak{z}] = \int \mathD \lambda \mathD a^\mu  \mathcal{D} \vb*{z}\, \exp\left(- \frac{1}{g} \int \dd^d \vb{x} \dd\tau \left[ \abs{D^\mu[a]\vb*{z}}^2 + \lambda(\abs{\vb*{z}}^2 - 1) \right] \right. \\
    + \left. \int \dd^d \vb{x} \dd\tau\, \mathfrak{z}^a(\vb{x}) z^*_\alpha(\vb{x},\tau) \gamma^a_{\alpha\beta} z_\beta(\vb{x},\tau) \right),
\end{split}
\end{equation}
where $\gamma^a$ are the generators of $SU(N)$, $\mathfrak{z}^a(\vb{x})$ is a real $(N^2-1)$-component static random field distributed according to
\begin{equation}
    \overline{\frakz^a(\vb{x})} = 0, \kern3em \overline{\frakz^a(\vb{x}) \frakz^b(\vb{y})} = \eta^2 \delta^{ab} \delta^{(d)}(\vb{x} - \vb{y}),
\end{equation}
$\eta^2$ is the variance of the disorder and overlines denote averages over configurations of the disorder with respect to this distribution. Using the replica trick to perform the disorder average, we obtain
\begin{align}
    \overline{\mathZ^n} &= \int \mathD \frakz \exp\left( - \int \dd^d \vb{x} \frac{\frakz^2}{2\eta^2} \right) \mathZ[\frakz]^n \nn \\
\begin{split}
    &= \int \mathD \lambda_j \mathD a^\mu_j \mathD \vb*{z}_j \, \exp\left(- \sum_{j = 1}^n \int \dd^d \vb{x} \dd\tau \left[  \abs{D_\mu[a] \vb*{z}_j}^2 + \lambda_j \left( \abs{\vb*{z}_j}^2 - \frac{1}{g} \right) \right] \right. \\
    &\kern12em +\left. \frac{N \eta^2}{2}\sum_{i,j=1}^n \int \dd^d \vb{x} \dd\tau_1 \dd\tau_2  \vert \vb*{z}_i^*(\tau_1) \cdot \vb*{z}_j(\tau_2) \vert^2 \right),
\end{split}
\end{align}
where we have used the identity $\gamma^a_{\alpha\beta} \gamma^a_{\gamma \delta} =  N \delta_{\alpha \delta} \delta_{\beta\gamma} - \delta_{\alpha\beta} \delta_{\gamma\delta}$, and the fact that $\abs{\vb*{z}(\vb{x},\tau)}^2 = 1$. Unlike in the classical theory studied in Ref. \cite{OBrien2024}, the quenched disorder generates bilocal in time interactions. However, the large-$N$ technique is still well-adapted to solving this problem (see, for example, Ref. \cite{Sachdev1993,Cugliandolo1998,Scammell2020}). The Hubbard-Stratonovich decoupling must be in the $SU(N)$ color singlet channel to apply the large-$N$ technique, so we obtain, after integrating out the $\vb*{z}_j$ fields,
\begin{align}
\begin{split}
    S_{\mathrm{eff}}/N = \Tr \ln\left( \delta(\tau_1 - \tau_2) \mathrm{diag}_r\left[- D_\mu^2[a] + \lambda\right] - \hat{\kappa} \right) - \sum_{j=1}^n \int \dd^d \vb{x} \dd\tau \frac{\lambda_j}{g_0}  \\
    + \sum_{i,j = 1}^n \int \dd^d \vb{x} \dd\tau_1\dd\tau_2 \frac{\kappa_{ij}(\tau_1,\tau_2) \kappa_{ji}(\tau_2,\tau_1)}{2\eta_0^2} ,
\end{split}
\end{align}
where $g = g_0/N$, $\eta = \eta_0/g_0$, $\mathrm{diag}_r(\,\cdot\,)$ denotes a matrix which is diagonal in replica indices, $\hat{\kappa}$ is the matrix with elements $\kappa_{ij}$, and $\Tr$ includes a trace over functional configurations and replica indices. The disorder Hubbard-Stratonovich field satisfies $[\kappa_{ij}(\tau_1,\tau_2)]^* = \kappa_{ji}(\tau_2,\tau_1)$, $\kappa_{ii}(\tau,\tau) = 0$, and transforms as a tensor under the local $U(1)$ symmetry $\vb*{z}_j(x) \rightarrow e^{i \phi_j(x)}\vb*{z}_j(x)$, $a_j^\mu(x) \rightarrow a_j^\mu(x) - \del^\mu \phi_j(x)$, $\kappa_{jk}(\vb{x},\tau_1,\tau_2) \rightarrow e^{i(\phi_j(x_1) - \phi_k(x_2))}\kappa_{jk}(\vb{x},\tau_1,\tau_2)$. To find the $N=\infty$ effective potential, we make the replica-symmetric ansatz $a^\mu_j=0$, $\lambda_j = m^2$, $\kappa_{ij}(\tau_1,\tau_2) = \kappa_0 (\tau_1 - \tau_2) -\delta_{ij} \kappa_1(\tau_1 - \tau_2) $. This yields the partition function
\begin{subequations}
\begin{align}
    \overline{\mathZ^n} &= e^{-N \mathcal{V} \overline{U_{\mathrm{eff}}^n}}, \\
\begin{split}
    \overline{U_{\mathrm{eff}}^n} &= \int \frac{\dd^d \vb{q}}{(2\pi)^d} \frac{\dd \omega}{2\pi} \ln \det \left( [\omega^2 + \vb{q}^2 + m^2 + \tilde{\kappa}_1(\omega)] \hat{I} - \tilde{\kappa}_0(\omega) \hat{X} \right) \\
    & - n \frac{m^2}{g_0} + n\int \dd\tau \frac{[\kappa_1(\tau) - \kappa_0(\tau)]^2}{2\eta_0^2} + n(n-1) \int \dd\tau \frac{[\kappa_0(\tau)]^2}{2\eta_0^2} + n \frac{\alpha}{\eta_0^2} [\kappa_0(0) - \kappa_1(0)],
\end{split}
\end{align}
\end{subequations}
where $\tilde{\kappa}_a(\omega)$ is the Fourier transform of $\kappa_a(\tau)$, $\alpha$ is the Lagrange multiplier which imposes the constraint $\kappa_{ii}(\tau,\tau) = 0$, $\hat{I}$ is the identity matrix in replica space and $\hat{X}$ is the matrix of all ones. In the limit $n\rightarrow 0$, we obtain the disorder-averaged effective potential
\begin{align}
\begin{split}
    \overline{U_{\mathrm{eff}}} = \int \frac{\dd^d \vb{q}}{(2\pi)^d} \frac{\dd \omega}{2\pi} \left[ \ln \left(\omega^2 + \vb{q}^2 + m^2 + \tilde{\kappa}_1(\omega)\right) - \frac{\tilde{\kappa}_0(\omega)}{\omega^2 + \vb{q}^2 + m^2 + \tilde{\kappa}_1(\omega)} \right]  \\
    -\frac{m^2}{g_0} +\int \dd\tau \frac{[\kappa_1(\tau)]^2 - 2 \kappa_1(\tau) \kappa_0(\tau)}{2\eta_0^2} + \frac{\alpha}{\eta_0^2}[\kappa_0(0) - \kappa_1(0)],
\end{split}
\end{align}
with the corresponding saddle point equations
\begin{subequations}
\begin{align}
    &\int \frac{\dd^d \vb{q}}{(2\pi)^d} \frac{\dd \omega}{2\pi}  \left[ \frac{1}{\omega^2 + \vb{q}^2 + m^2 + \tilde{\kappa}_1(\omega)} + \frac{\tilde{\kappa}_0(\omega)}{[\omega^2 + \vb{q}^2 + m^2 + \tilde{\kappa}_1(\omega)]^2} \right] = \frac{1}{g_0}, \\
    &\int \frac{\dd^d \vb{q}}{(2\pi)^d} \left[ \frac{1}{\omega^2 + \vb{q}^2 + m^2 + \tilde{\kappa}_1(\omega)} + \frac{\tilde{\kappa}_0(\omega)}{[\omega^2 + \vb{q}^2 + m^2 + \tilde{\kappa}_1(\omega)]^2} \right] = \frac{\alpha - \tilde{\kappa}_1(\omega) + \tilde{\kappa}_0(\omega)}{\eta_0^2}, \\
    &\int \frac{\dd^d \vb{q}}{(2\pi)^d} \frac{1}{\omega^2 + \vb{q}^2 + m^2 + \tilde{\kappa}_1(\omega)}  = \frac{\alpha - \tilde{\kappa}_1(\omega)}{\eta_0^2}, \\
    &\kappa_0(0) - \kappa_1(0) = 0.
\end{align}
\end{subequations}
The Lagrange multiplier can be eliminated from the second and third equations to obtain the equivalent expressions
\begin{subequations}
\begin{align}
    \int \frac{\dd^d \vb{q}}{(2\pi)^d} \frac{\tilde{\kappa}_0(\omega)}{[\omega^2 + \vb{q}^2 + M^2 + \tilde{\kappa}_1(\omega) - \tilde{\kappa}_1(0)]^2}  = \frac{\tilde{\kappa}_0(\omega)}{\eta_0^2}, \\
    \int \frac{\dd^d \vb{q}}{(2\pi)^d} \left[ \frac{1}{\omega^2 + \vb{q}^2 + M^2 + \tilde{\kappa}_1(\omega) - \tilde{\kappa}_1(0)} - \frac{1}{\vb{q}^2 + M^2} \right] = -\frac{\tilde{\kappa}_1(\omega) - \tilde{\kappa}_1(0)}{\eta_0^2},
\end{align}
\end{subequations}
where $M^2 = m^2 + \tilde{\kappa}_1(0)$ is the physical gap. It is simple to see from the saddle point equations that $\tilde{\kappa}_1(0)$ must actually be infinite, but eliminating the Lagrange multiplier regularizes the equations. Since $\eta_0$ is independent of frequency, we see that $\tilde{\kappa}_0(\omega)$ must have the form
\begin{equation}
    \tilde{\kappa}_0(\omega) = \kappa_0 (2\pi) \delta(\omega),
\end{equation}
which implies
\begin{equation}
    \frac{\kappa_0}{4\pi M^2} = \frac{\kappa_0}{\eta_0^2} ,
\end{equation}
and hence, $\kappa_0 = 0$ or $\kappa_0 \neq 0$ and $M^2 = \eta_0^2/4\pi$. Note that we must have $M^2 \geq \eta_0^2/4\pi$ as $\eta_0 \rightarrow 0$ since the clean theory can also be gapped. The second equation can be inverted exactly to obtain
\begin{equation}
    \tilde{\kappa}_1(\omega) - \tilde{\kappa}_1(0) = -\omega^2 - M^2 - \frac{\eta_0^2}{4\pi} W_{-1}\left(-\frac{4\pi M^2}{\eta_0^2} e^{-4\pi(\omega^2 + M^2)/\eta_0^2}\right),
\end{equation}
where $W_k(z)$ is the $k$\textsuperscript{th} branch of the product logarithm function; i.e., the inverse function of $z = W e^W$ \cite{Corless1996}. The $k=-1$ branch is selected to satisfy the initial condition at $\omega = 0$. As $\abs{\omega} \rightarrow \infty$, $\tilde{\kappa}_1(\omega) - \tilde{\kappa}_1(0) \simeq (\eta_0^2/4\pi)\ln(\omega^2/M^2)$, while the behavior as $\abs{\omega} \rightarrow 0$ depends on whether $\eta_0^2 < 4\pi M^2$ or $\eta_0^2 = 4\pi M^2$,
\begin{subequations}
\begin{align}
    \tilde{\kappa}_1(\omega) - \tilde{\kappa}_1(0) \simeq \left(\frac{4\pi M^2}{\eta_0^2} - 1 \right)^{-1} \omega^2 + \mathcal{O}(\omega^4), \kern3em \eta_0^2 < 4\pi M^2,\\
    \tilde{\kappa}_1(\omega) - \tilde{\kappa}_1(0) \simeq \frac{\eta_0}{\sqrt{2\pi}} \abs{\omega} - \frac{1}{3} \omega^2 + \mathcal{O}(\abs{\omega}^3), \kern3em \eta_0^2 = 4\pi M^2.
\end{align}
\end{subequations}
Given these observations, the first saddle point equation can be made finite with the usual coupling constant renormalization
\begin{equation}
    \frac{1}{g_0} = \frac{1}{g_R} \left(1 + g_R  \int^\Lambda \frac{\dd^d \vb{q}}{(2\pi)^d} \frac{\dd \omega}{2\pi} \frac{1}{\omega^2 + \vb{q}^2 + \mu^2} \right),
\end{equation}
so that
\begin{equation}
    \frac{1}{4\pi}\int \frac{\dd \omega}{2\pi} \ln \left(-\frac{\eta_0^2}{4\pi\omega^2} W_{-1}\left(-\frac{4\pi M^2}{\eta_0^2} e^{-4\pi(\omega^2 + M^2)/\eta_0^2}\right) \right) = \frac{\kappa_0}{\eta_0^2} - \frac{1}{g_R} + \frac{\mu}{4\pi}.
\end{equation}
When $\kappa_0 = 0$, this is an implicit equation for $M^2$, while for $\kappa_0 \neq 0$ we have $M^2 = \eta_0^2/4\pi$ and the saddle point equation yields $\kappa_0(\eta_0)$. In both cases, the integral must be evaluated numerically, which is challenging to do precisely while in its current form. Instead, the change of variables
\begin{equation}
    w e^{-w} = \frac{4\pi M^2}{\eta_0^2} e^{-4\pi(\omega^2 + M^2)/\eta_0^2},
\end{equation}
allows us to re-write the integral as
\begin{figure}[!t]
    \centering
    \includegraphics[scale=0.5]{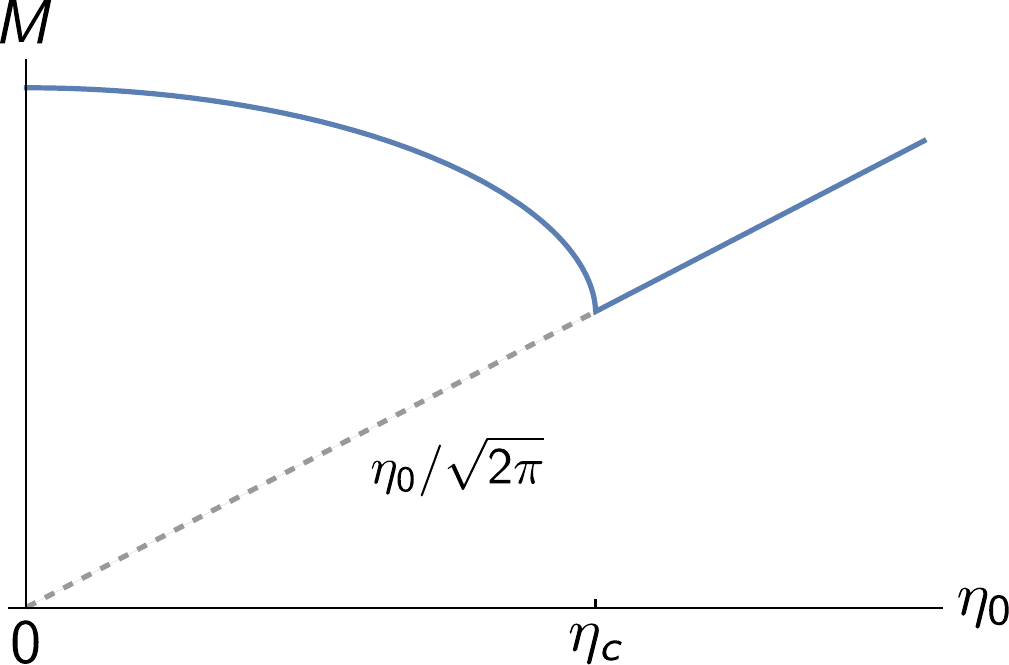}
    \caption{The physical gap $M$ as a function of disorder strength $\eta_0$ for $g_R > g_c = 4\pi/\mu$. For $\eta_0 < \eta_c$, the mass is strictly greater than its lower bound $\eta_0/\sqrt{2\pi}$ (shown as a dashed line), and tends towards its clean value as $\eta_0 \rightarrow 0$. At weak disorder, the mass can be seen empirically to have the power-law behavior $M(\eta_0) \simeq m_0(1 - \eta_0^2/\eta_c^2)^{1/4}$ where $m_0 = \mu(1-g_c/g_R)$.}
    \label{fig:cpn_gap}
\end{figure}
\begin{equation}
    \frac{\eta_0}{2\pi (4\pi)^{3/2}} \int_0^\infty \dd w \ln\left( \frac{w + \alpha}{w-\ln(1 + w/\alpha)} \right) \frac{1-1/(w+\alpha)}{\sqrt{w - \ln (1 + w/\alpha)}} = \frac{\kappa_0}{\eta_0^2} - \frac{1}{g_R} + \frac{\mu}{4\pi},
\end{equation}
where $\alpha = 4\pi M^2/\eta_0^2$. While this expression is not necessarily any more instructive, it is vastly simpler to evaluate numerically. The mass $M$ as a function of $\eta_0$ for $g_R > g_c =  4\pi/\mu$ is shown in Fig. \ref{fig:cpn_gap}. In the regime where $M^2 = \eta_0^2/4\pi$, we evaluate the integral to find
\begin{equation}
    \frac{\kappa_0}{\eta_0^2} \approx \frac{1}{g_R} - \frac{1}{g_c} + 0.03941 \times  \eta_0 .
\end{equation}
Since we must have $\kappa_0 \geq 0$ for the action to be positive definite, this defines the critical disorder strength
\begin{subequations}
\begin{align}
    \eta_{c} = \begin{cases}
        0, \kern3em & g_R \leq g_{c},\\
        2.019 \times \mu \left(1 - g_c/g_R \right), \kern3em & g_R > g_{c} ,
    \end{cases}
\end{align}
\end{subequations}
This implies that for $g_R \leq g_c$, the system is always strongly disordered; any infinitesimal amount of disorder is enough to produce a condensate of the static response strength $\kappa_0$. This is physically intuitive, since the gapless $SU(N)$ Goldstone modes in the broken symmetry phase of the clean $\mathbb{C}P^N$ model are much more strongly affected by disorder than the gapped modes in the symmetric phase.

\bibliography{refs}

\end{document}